\documentclass[11pt]{article}
\usepackage{amssymb,amsthm,amsfonts,psfig,epsfig,axodraw,color}
\def\gappeq{\mathrel{\rlap {\raise.5ex\hbox{$>$}} {\lower.5ex\hbox{$\sim$}}}}
\def\lappeq{\mathrel{\rlap{\raise.5ex\hbox{$<$}} {\lower.5ex\hbox{$\sim$}}}}
\def\beq{\begin{equation}} 
\def\eeq{\end{equation}} 
\def\bea{\begin{eqnarray}}
\def\eea{\end{eqnarray}}
\def\bq{\begin{quote}} 
\def\eq{\end{quote}}

\def\nn{\nonumber}
\def\ep{\epsilon}

\def\ti{\tilde}
\def\Md{\bar M}

\def\ie{{i.e.}}
\def\eg{{e.g.}}

\def\viceversa{{vice versa}}

\begin{document} 
\pagestyle{empty}  
\begin{flushright}  SACLAY-T02/015
\end{flushright}  
\vskip 2cm    

\def\thefootnote{\fnsymbol{footnote}}

\begin{center}  
{\large \bf LARGE SOLAR ANGLE AND SEESAW MECHANISM: \\
A BOTTOM-UP PERSPECTIVE}
\vspace*{5mm}    
\end{center}  

\vspace*{5mm} \noindent  
\vskip 0.5cm  
\centerline{\bf St\'ephane Lavignac, Isabella Masina and Carlos A. Savoy 
\protect\footnote{E-mails: lavignac@spht.saclay.cea.fr, masina@spht.saclay.cea.fr, 
savoy@spht.saclay.cea.fr.}}

\vskip 1cm
\centerline{\em Service de Physique Th\'eorique \protect\footnote{Laboratoire de la Direction 
des Sciences de la Mati\`ere du Commissariat \`a l'\'Energie Atomique et Unit\'e de Recherche 
Associ\'ee au CNRS (URA 2306).} , CEA-Saclay}

\centerline{\em F-91191 Gif-sur-Yvette, France} 
\vskip2cm 
  
\centerline{\bf Abstract}  

In addition to the well established large atmospheric angle, a large solar angle
is probably present in the leptonic sector. 
In the context of the see-saw and by means of a bottom-up approach, 
we explore which patterns for the Dirac and Majorana right-handed mass matrices  
provide two large mixings in a robust way and with the minimal amount of tuning. 
Three favourite patterns emerge, which have a suggestive physical interpretation in terms of 
the role played by right-handed neutrinos: in both solar and atmospheric sectors,
either a single or a pseudo-Dirac pair of right-handed neutrinos dominates.
Each pattern gives rise to specific relations among the 
neutrino mixing angles and mass differences, which lead to
testable constraints on $U_{e3}$.
The connection with the rate of LFV charged lepton decays is also addressed. 
\vspace*{1cm} 
\vskip .3cm  \vfill   \eject 

\newpage  
\setcounter{page}{1} \pagestyle{plain}


\def\thefootnote{\arabic{footnote}}
\setcounter{footnote}{0}

\section{Introduction}

Combined results \cite{fit_including_SNO} from SNO \cite{SNO} and Super-Kamiokande \cite{solar} suggest
that another large mixing angle is present in the leptonic sector,
in addition to the well established large mixing angle involved in atmospheric oscillations
\cite{atmlarge}.
CHOOZ data \cite{CHOOZ}, on the contrary, constrain the third angle $\theta_{13}$
to be smaller than the Cabibbo angle.
Such an interestingly different pattern of lepton mixings as compared to quark mixings 
should be interpreted as the result of some specific property characterizing the 
physics beyond the Standard Model.  

The see-saw mechanism \cite{seesaw} has emerged as the most elegant explanation of the smallness
of neutrino masses: $ {\cal M}_\nu =Y^T M^{-1} Y v^2 $, $Y v$ and $M$ being respectively the 
Dirac and the Majorana mass matrices
\footnote{We are dealing here with the simplest version of the see-saw mechanism, obtained by just adding 
three right-handed neutrinos to the matter content of the Standard Model.}. 
Triggered by such impressive experimental results, 
one would like to understand if the requirement of a large solar angle leads  
to some specific realization of the see-saw mechanisms. 
Then, one could look for
theoretical implications - like those on the underlying flavour symmetry - and also for
phenomenological implications - like the magnitude of $U_{e3}$ and
the rate of lepton flavour violating (LFV) charged lepton decays \cite{tau_mu_gamma, mu_e_gamma}.
Many see-saw neutrino mass models have been proposed in the last years
(for a review and a set of references see e.g. \cite{isaIJMPA}), in particular
in connection with flavour symmetries and grand unified theories. From the model building point 
of view, with an hierarchy between 
$m_@ \equiv \sqrt{\Delta m^2_{atm}}$ and $m_{\odot} \equiv \sqrt{\Delta m^2_{\odot}}$
it is much more challenging to end up with a large mixing than with a small one. 
This holds for the atmospheric angle, so that it holds a fortiori when requiring a large solar 
angle too. Indeed, either one imposes various tunings between the many elements of $Y$ and $M$ 
\cite{Irges98}, or one has to deal with apparently intricate patterns for $Y$ and $M$ which could 
only be obtained from highly non-trivial flavour symmetries \cite{Smir95, King99, afm1, bizarre, u2}.  

To overcome the unavoidably fragmentary picture which arises by considering 
the many specific models \footnote{For previous interesting attempts see \cite{prev}.}, 
one should investigate in a model independent way 
which essential features of $Y$ and $M$ can account for two large angles and two mass scales.
In this paper we tackle this problem from a bottom-up point of view 
and we identify those $Y$ and $M$ patterns which 
give two large mixings because of their own {\it structure} with the minimal amount of tuning
\footnote{We will not consider the extended flavour democracy scenario (see e.g. \cite{efd}
and references therein).}.
Since their structure is just the reflection of the different roles played by right-handed neutrinos, 
from the analysis of such 'natural' patterns it will turn out that few suggestive
possibilities for such roles. As it will be discussed, interesting
phenomenological implications also follow, in particular for the magnitude of $U_{e3}$.
 
At low energy and in the basis were charged leptons are diagonal,
the neutrino sector is described in terms of nine physical quantities, while
at high energy, that is above the scale where right handed neutrinos are integrated out,
it is specified by eighteen. Casas and Ibarra \cite{Casas01} identify
the {\it hidden parameters} of the see-saw with
the six parameters of a complex orthogonal matrix acting on the left side of $L \equiv
M^{-1/2} Y v$ and the three right handed neutrino mass eigenvalues.
Of course, it is unrealistic at present to follow a complete bottom-up approach
to reconstruct $Y$ and $M$ \cite{Davidson01}.
In fact, even if other observable quantities, like the rate of LFV processes 
\cite{Casas01, athand, lms1}, 
are sensitive to the hidden see-saw parameters (see e. g. \cite{Davidson01, sens}), 
such observables -- even if measured -- provide upper limits, so that one is still far from having any 
direct access to the elements of $Y$ and $M$.

In order to fill this gap between the number of observables at low and high energy,
one needs some 'theoretical salt'. 
The construction of specific models \cite{more} according to the rules dictated by some 
flavour and/or GUT symmetry is a top-down approach to the problem of neutrino masses and mixings
and the only ambiguities are related to the predictive power of the theory.
In this sense, the neutrino oscillation data can test the different theoretical high energy models,
but of course they cannot select a particular one. 

In the bottom-up approach of this paper, 
we adopt instead the requirements of robustness and economy as additional ingredients.
Indeed, our aim is to isolate those $\{Y, M\}$ which account
for neutrino mixings and the hierarchy $m^2_@ >> m^2_{\odot}$ in a natural way.
In this spirit, we look for the structures of $Y$ and $M$ which are stable, in the sense that small
perturbations in the parameters just induce small corrections in the physical observables. 
Thus, a large mixing angle cannot result from a tuning of the many elements of $Y$ and $M$,
which would be presumably unstable, but rather from a simple relation between a couple of 
matrix elements.
As another consequence, we exclude the nearly degenerate neutrino mass spectrum because the mixing
angles are very unstable under radiative corrections \cite{RGE_nu}.
The approach is also economical as it minimizes the number of the conditions on the parameters,
while allowing most of the parameters to remain generic. 
One can derive some generic predictions in the framework of each pattern
in the form of quantitative constraints between experimental observables.
In this sense, we may decide if one generic pattern is plausible or disfavoured by present and future 
data.
We carry out our analysis for the two viable scenarios for the neutrino mass spectrum,
namely the hierarchical (Hi) and inverse hierarchical (iH) case. 

We first apply the stability criteria to the analysis of the atmospheric angle. 
In Section 2, we display the patterns of $\cal{M}_\nu$ consistent with 
large atmospheric mixing and $m^2_@ >> m^2_{\odot}$ for these two scenarios.
Then, we discuss how to obtain them with the least amount of tuning. 
In the (Hi) case, the naturalness criteria for the atmospheric sector 
are satisfied by the well known {\it dominance} hypothesis \cite{Smir93, King99, afm1}: 
the mass scale of the heaviest light neutrino is essentially provided by only one right 
handed neutrino, and the fact that the angle is large is obtained
by imposing that its Yukawa couplings to the $\mu$ and $\tau$ have comparable strength.
The dominance hypothesis has interesting phenomenological \cite{lms1, HKlept} and theoretical \cite{afm1} 
consequences. 
Also for the (iH) case, as will be shown, it is possible to recognize a dominance mechanism
which would be solidly realized by a pseudo-Dirac pair of right-handed neutrinos having a suitable 
hierarchy in their Yukawa couplings. 
Models for (iH) have been studied in particular in association with the breaking of 
$L_e-L_\mu-L_\tau$ \cite{pet82, barb98, iHcase, iHbreak}.

Along this line of reasoning, the next step is to investigate which pattern 
of the sub-leading elements of ${\cal M}_\nu$ accounts for the presence of a large or maximal solar angle 
in the most natural way. This is done in Section 3.
Dealing with sub-leading contributions is a delicate task in the 
presence of several small parameters. Moreover, one has to take into account the CP violating phases 
\cite{cpbranco} potentially present in $Y$. We carry out this program in a quantitative and systematic 
way. Of course some results are
already known as empirical rules from many previous analyses, in particular
from the model building point of view (see, e.g. \cite{Smir93, Smir95, King99, prev}). 
We perform the calculation in a systematic way 
to see explicitly how much these intuitive ideas are reliable and if there are important
exceptions. Fortunately, a more quantitative reappraisal is already possible with
reliable approximate expressions, where to recognize immediately the leading and the sub-leading 
contributions, keeping under control the order of magnitude of the negligible sub-subleading contributions.
These expressions show that a large solar angle represents a strong constraint and suggest
peculiar structures for $Y$ and $M$. This also explains why it is non trivial to build models which account 
for two large angles. It turns out that the quantitative
analysis supports patterns which have a remarkably simple physical
interpretation in terms of how right-handed neutrinos couple to each light neutrino
flavour eigenstate.
We then describe in some detail these suggestive patterns. 
In the (Hi) case, as we have already mentioned, the atmospheric mass scale and mixing angle
essentially arise from just one of the right-handed companions, so that the solar sector
is directly linked to the properties of the other two right-handed neutrinos. 
Two basic possibilities automatically lead to a sufficiently 
large solar angle which we will refer to as {\it double dominance} and {\it pseudo-Dirac} 
for short. 
The first case presents also in the solar sector a mechanism analogous to the one at work for the 
atmospheric sector:
one among the remaining two right-handed neutrinos is
essentially responsible for the solar mass scale and the large solar angle is obtained
when it couples with the same strength to the $e$ and to the $\mu$ and $\tau$ combination 
orthogonal to the heaviest neutrino eigenstate. 
To approach a maximal mixing would need a precise equality of the couplings.
In the double dominance each light neutrino has its preferred right-handed companion.
The second possibility for having a stable large solar angle, 
arises when the solar sector of the light neutrino mass matrix
possesses a pseudo-Dirac structure. The solar angle approaches its maximal value
when the Dirac structure is stressed.
This naturally arises when 
the two remaining right-handed neutrinos possess a pseudo-Dirac structure and a suitable hierarchy
of Yukawa couplings. 
In the (iH) case, the mass of the pseudo-Dirac pair is close to the atmospheric scale;
$m_\odot$ being much smaller, the pseudo-Dirac structure has to be
well approached.
As a consequence, the deviation of the solar angle from $\pi/4$ has to be very tiny.
The relation is so strong that only the LOW solution is really viable. 

From the point of view of introducing the minimal amount of tuning, these three scenarios are
preferred. They are all characterized by the dominance of either a single or a pseudo-Dirac
pair of right-handed neutrinos in both the solar and atmospheric sectors. 
On the contrary, the simplest flavour
symmetry, a plain $U(1)$ symmetry \cite{FN} with all charge assignments of the same sign, would predict
a democratic dominance pattern, with all right handed neutrinos contributing roughly the
same to all light neutrino masses.
Then, the above considerations show that the most natural structures from a bottom-up
perspective are realized only by means of 
non-trivial flavour symmetries such as $U(1)$'s with holomorphic zeros from superymmetry
\cite{afm1, bizarre} or non-abelian flavour symmetries \cite{u2}. Alternatively, one could 
consider the above three possibilities as recipes for model building.
Of course, deviations from these structure require introducing  more 
correlations among the parameters.

The above natural patterns for $Y$ and $M$ are also interesting from a
phenomenological point of view. Being quite constrained, 
it is straightforward to look for their predictions.
In particular, those for $U_{e3}$ are remarkably different:
in the (Hi) case, $U_{e3}$ could not escape detection if the experimental sensitivy reaches few percent,
while in the (iH) case, on the contrary, only LOW is viable and $U_{e3}$ is predicted to be smaller than 
one per mille. 
All the three scenarios considered predict a rate for $\beta \beta 0 \nu$ too small to be observed.
In Section 4 we discuss the connections with the LFV decays $\mu \rightarrow e \gamma$ and
$\tau \rightarrow \mu \gamma$, emphasising the strong dependence
on the absolute scales of right-handed neutrino masses and on the other hidden see-saw parameters. 
On the contrary, sizeable electron and muon electric dipoles 
would provide an indication for a hierarchy in the right-handed neutrino spectrum \cite{dedmu}.

In section 5 we show that the hidden parameters identified
with the complex orthogonal matrix admit an interesting physical interpretation
as a dominance matrix. Outlook and conclusions are presented in Section 6.

\vspace*{2cm}

\section{Accounting for the Atmospheric Neutrino Mixing Oscillations}

\subsection{The See-Saw Parameters}\label{sec:seesaw}

In the see-saw model \cite{seesaw} the effective neutrino mass matrix is given in
terms of the Dirac mass matrix $Yv$ and the Majorana mass  matrix $M$
for right-handed neutrinos as
\beq
{\cal M}_\nu  =  U^* \left( \matrix{ m_1 & & \cr & m_2 & \cr & & m_3} 
\right) U^{\dagger} =  Y^T \frac{v^2}{M} Y\ ,
\eeq 
where $Y$ is defined in the basis where the charged leptons are the
mass eigenstates ($e, \mu, \tau $) and, for definiteness, we choose the
basis of $M$ eigenstates $(M_x,M_y,M_z >0)$ for the right-handed
neutrinos. Their ordering will be defined below. The complete
determination of $\cal{M}_\nu$ would provide us with 9 parameters: 3
eigenvalues of $\cal{M}_\nu$, 3 angles and 3 phases of $U$. Instead,
the see-saw model has 21 parameters: 18 complex numbers of the matrix
$Y$ and 3 eigenvalues of $M$. At the lepton flavour violation scale,
in the framework of the see-saw model (\ie , the SM or the MSSM plus
the right-handed neutrino masses), 3 phases in $Y$ can be
reabsorbed in the definition of the leptonic fields. There are 18-9=9
more parameters in the see-saw model than those to be measured in the
matrix  $\cal{M}_\nu$.

Let us define the matrix
\beq
L \equiv M^{-1/2}  Y v =\left( \matrix{ x_1 & x_2 & x_3 \cr 
y_1 & y_2 & y_3 \cr  
z_1 & z_2 & z_3  }\right)\ ,
\label{laL}
\eeq
such that ${\cal M}_\nu=L^T L$ and the bases are defined as above (3
phases in the 9 complex matrix elements are unphysical at low
energies). The see-saw mechanism involves $L$ and the $M$ eigenvalues.
If CP is conserved in the lepton sector of the whole theory (i.e. at energies
above the scale where right-handed neutrinos decouple), the phases of
the fields can be chosen to have $Y$, $M$ and ${\cal M}_\nu$ to be real
matrices. Negative
$M$ (${\cal M}_\nu$) eigenvalues correspond to CP odd right-handed 
(left-handed) neutrino eigenstates. It is worth stressing the 
following property: $M$ and ${\cal M}_\nu$ possess the same number of 
negative eigenvalues. Actually, we shall prefer here the convention where 
$M$ and ${\cal M}_\nu$ eigenvalues are all positive, and multiply the 
lines in $Y$ and $L$ that correspond to a CP odd right-handed neutrino
state by a phase $i.$ 
In general however, the phases may be important and the $L$ matrix elements
are complex, while $M_x$, $M_y$, $M_z$ are positive.   

As noticed by Casas and Ibarra \cite{Casas01}, extracting $L$ from
$\cal{M}_\nu$ suffers from the ambiguity
\beq
{\cal M}_\nu=L^T L=(R L)^T (R L)\ ,
\label{laR}
\eeq  
where $R$ is any complex orthogonal matrix, i.e., any element of
$O(3,\mathbb{C})$. More explicitly, one has
\beq
{\cal{M}_\nu}_{ij}=x_i x_j +y_i y_j +z_i z_j , 
\eeq
which is clearly invariant under
\beq
\left(  \matrix{ x_i \cr y_i \cr z_i } \right) \longrightarrow R 
\left(  \matrix{ x_i \cr y_i \cr z_i } \right) ,
~~~~~i=1,2,3\ ,
\eeq
where $R$ can be parameterized as
\beq
R=e^{\alpha J_z} e^{\beta J_x} e^{\gamma J_z} ,
\label{rot}
\eeq 
with $\alpha, \beta, \gamma$ three complex numbers and $J_{x,y,z}$ are the
$O(3)$ generators. Therefore the solutions $L$ of eq. (\ref{laR}) can be
written as
\beq
L= \ti{R} \left( \matrix{ 0 & 0 & \ti{x}_3 \cr 
0 & \ti{y}_2 & \ti{y}_3 \cr  
\ti{z}_1 & \ti{z}_2 & \ti{z}_3 } \right)\ ,
\label{laLbis}
\eeq 
where $\ti{R} \in O(3,\mathbb{C})$ remains arbitrary and the matrix elements
$\ti{x}_3,..., \ti{z}_3$ (among which 3 are real, the other 3 being complex)
can be determined from $\cal{M}_\nu$, up to sign ambiguities. Of course, both
$\ti{R}$ and the eigenvalues of $M$ are physical in the framework of the see-saw
model, but other experiments are needed to access them, as we also
discuss below.

In some cases discussed below, the matrix $R$ defined by Casas and Ibarra
(which should not be confused with $\ti{R}$) can be interpreted as a
dominance matrix, which associates each light neutrino mass eigenvalue
$m_i$ with a given combination of right-handed neutrino masses. In general
however, its interpretation is not transparent because it is not
unitary. For this reason, we prefer to work with (\ref{laL}) rather
than with (\ref{laLbis}) and (\ref{rot}).

\subsection{See-Saw Realizations of Mass Hierarchies and Large Mixings}
\label{sec:mechanisms}

Our aim in this section is to identify the patterns for the see-saw
parameters -- more exactly, the patterns for $L$, since we do not assume
any a priori hierarchy for the right-handed neutrino masses -- that fit
in the most natural way the experimental results. Because of
the surplus in the parameters, one must adopt a simplifying assumption
and it looks reasonable to assume each feature of $\cal{M}_\nu$ to be
due to a simple relationship in $L$, such as two parameters being 
almost equal or very different, rather than to assume an elaborate
conspiracy between the many elements of a matrix. We also apply the same
considerations in linking the properties of $L$ to those of $Y$ and $M$.
We first sketch the two patterns that allow to minimize the tuning
of the seesaw parameters. 

The first feature is the hierarchy between $m^2_{\odot}$ and
$m^2_{@}$. A way to account for one eigenvalue being
much larger than another, is to assume a dominance mechanism
\cite{Smir93, King99, afm1}, in which one right-handed neutrino, associated to a line in
$L$ which we take to be $z$, gives the dominant contribution to the
larger eigenvalue $m_i$ of $\cal{M}_\nu$. Then
\beq
m_i = \sum_j {\left( (LU)_{ji} \right) }^2 
\approx {\left( \sum_k U_{ki}z_k \right)}^2 
\approx \sum_k {\left( z_k \right)}^2 \equiv z^2\ ,
\eeq
and (phases are unimportant here) the corresponding eigenstate is
approximately $z_j\nu ^j /z,$ so that 
\beq
U_{ei} \sim z_1 /z , \ \ \ \ U_{\mu i}\sim z_2 /z ,
\ \ \ \ U_{\tau i} \sim z_3 /z ~~. \label{compon}
\eeq
Within the dominance assumption, ${\cal M}_{\nu ij} $ is given by $z_i
z_j$ plus corrections of the order of the smaller eigenvalues. Whenever
some $z_i / z$ becomes small as compared to these corrections the
corresponding expression in (\ref{compon}) has to be corrected. Thus,
the larger eigenvalue of $\cal{M}_\nu$ is here due to a right-handed
neutrino with some larger Yukawa couplings or a relatively small mass.

The second feature of neutrino oscillations is the favoured large
mixing angle solar solutions. A large mixing angle is naturally
obtained with a pseudo-Dirac structure of ${\cal M}_\nu$, i.e. when
the dominant entries are off-diagonal (as discussed below, this cannot
be applied to the atmospheric mixing angle if almost degenerate
neutrinos are excluded). In this case at least two mass eigenstates are
very close in mass, with CP-violating Majorana phases differing by
$\pi/2$ (opposite CP parities in the CP-conserving case), and strongly
mixed.  Such a structure requires strong correlations between the
couplings of at least two right-handed neutrinos. The most natural
origin for these correlations is the existence of a pseudo-Dirac pair
of right-handed neutrinos -- recall that, in the CP-conserving case,
there should be as many right-handed neutrinos with negative CP
parities as light neutrinos with negative CP parities. We concentrate
on the $2 \times 2$ pseudo-Dirac sector of the mass matrices.  Then
there is a basis in which $M$ takes the form:
\beq
\left(  \matrix{  0 & \bar{M} \cr 
\bar{M} & 0 } \right) \qquad \left( +\ \mbox{small corrections} \right)\ ,
\label{psDimass}
\eeq
where we have chosen the pseudo-Dirac pair of right-handed neutrinos to
correspond to two lines in $L$, \eg , $x$ and $y.$
Neglecting the smaller contributions (deviations from the pseudo-Dirac
pattern) one gets for the $L$ matrix elements:
\beq
y_i = \frac{({\bar{Y}}_{2i}+{\bar{Y}}_{1i})}{\sqrt{2\bar{M}}}\ v\ , \qquad 
x_i =\frac{ i({\bar{Y}}_{2i}-{\bar{Y}}_{1i})}{\sqrt{2\bar{M}}}\ v\ , 
\label{psDixy}
\eeq
where the Yukawa matrix ${\bar{Y}}$ is in the basis defined by
(\ref{psDimass}). Eq. (\ref{psDixy}) leads to a pseudo-Dirac structure
for ${\cal M}_\nu$, with $m_1 \simeq -m_2$, provided the hierarchy among
$\bar{Y}$ couplings is as follows:
\beq
{\bar{Y}}_{21} \gg {\bar{Y}}_{11} \qquad \mbox{and} \qquad 
{\bar{Y}}_{12}\gg {\bar{Y}}_{22} \qquad 
(\mbox{or}\ {\bar{Y}}_{2j}\rightleftharpoons {\bar{Y}}_{1j} ) \ .  
\label{psDiY}
\eeq
so that $|y_1 +ix_1| \ll |y_2 +ix_2|$ and $|y_2 -ix_2| \ll |y_1 -ix_1|$
or \viceversa .

The diagonalization of the resulting $\cal{M}_\nu$ yields a large
mixing angle which becomes closer to $\pi/4$ as the pseudo-Dirac
configuration is better approached. Therefore, a close to maximal mixing
angle in neutrino oscillations (as indicated by the solar data in the LOW
region) suggests a pseudo-Dirac structure in $M$ and in $\cal{M}_\nu$,
which requires in turn that some particular Yukawa couplings are
predominant. However this option cannot be advocated for the atmospheric
oscillations, because it is not possible to accommodate solar neutrino
oscillations within a 3-neutrino scheme with $m_2 \simeq - m_3 > m_@ .$

As already stressed, an almost degenerate mass spectrum -- perhaps the
most obvious explanation for the large mixing angles -- is not yet
excluded by experiments. But it is generically disfavoured because the
large mixing angles become strongly scale dependent, unless some
compensation mechanism is introduced \cite{RGE_nu}.

\subsection{Patterns for $\cal{M}_\nu$}\label{sec:patterns}

Once solutions with three almost degenerate eigenvalues are excluded as
unstable under the RGE evolution, the largest $m_i$ eigenvalue must be
${\cal O} (m_{@})$. Furthermore, the matrix $\cal{M}_\nu$ must be consistent
with the experimental facts:\\ 
a) $m^2_{\odot} = m^2_2-m^2_1 \ll m^2_@$,
namely the hierarchy between solar and atmospheric neutrino mass
differences;\\ 
b) large atmospheric mixing angle $\theta_{23} \approx \pi/4$ and good
evidence for large solar mixing angle $\theta_{12} \approx \pi/4$ as well;\\ 
c) the CHOOZ constraint $|U_{e3}| < 0.2$.

There are basically two patterns for $\cal{M}_\nu$ that are consistent
with these properties (the choice of the ordering $i=1,2,3$ of the
$m_i$ is suitable for the standard conventions), which we now discuss
in turn.

\subsubsection{Hierarchical pattern: $m_3 \simeq m_@ \gg m_1, m_2$} 

The properties a), b), c) lead to a pattern for $\cal{M}_\nu$
consistent with the assumption of the dominance of the atmospheric
oscillations by one right-handed neutrino, namely (the intervals at 90\%
C.L. are taken from Ref. \cite{Gonz01}),
\beq
{{\cal M}_\nu}_{ij} = z_i z_j (1+ {\cal O} (\rho))
\eeq
with 
\bea
\sum_{i=1}^{3} z_i^2 \simeq m_@ = (4-8) \times 10^{-2}\, \mbox{eV} \nn\\
\frac{z_2}{z_3} \simeq \tan \theta_{23} = (0.6 - 1.7) \\
|z_1/z_3| \lesssim \tan \theta_{13} < 0.2 \nn
\eea
and ${\cal O} (\rho)$ denotes a matrix whose entries are at most of order
$\rho,$ with $\rho=m_2/m_3$. Thus this pattern is characterized by dominant
matrix elements of order $m_@$ in the $(2,3)$ submatrix of $\cal{M}_\nu$, so
that $z_2 \approx z_3 \approx \sqrt{m_@ /2}$, while the other are smaller since
$z_1 \lesssim \sqrt{m_@/2} \tan \theta_{13}$. This means that one right-handed
neutrino dominates the atmospheric oscillations. According to
the values of $x_i, y_i$, there are several possibilities to obtain
$m_\odot$ and $\theta_{12}$ from the other two right-handed neutrinos,
as discussed in the next section.

In a bottom-up approach, it is convenient to redefine $L$ by a rotation
\beq
\left( x_2, ~x_3 \right) \longrightarrow
\left( x_-, ~x_+ \right) = 
\left( \cos \theta_0 x_2 - \sin \theta_0 x_3,
~\cos \theta_0 x_3 + \sin \theta_0 x_2 \right)\ ,
\label{rotaz}
\eeq
where $\theta_0 = \arctan (z_2/z_3)$, and analogously for $(y_2, y_3)
\rightarrow (y_-, y_+)$,  $(z_2, z_3) \rightarrow (z_-, z_+)$. Since
$\theta_0 \simeq \theta_{23}$, the ($+$) components are approximately aligned
with the heaviest neutrino mass eigenstate $\nu_3,$ while the ($-$)
components are orthogonal. After this rotation one has $z_-=0$ and
$z_+ \simeq \sqrt{m_@}$, which implies
\beq
{\cal M}_{\nu +-}, ~~ {\cal M}_{\nu - -} \ll  {\cal M}_{\nu ++} \simeq m_@ ~.
\eeq
The atmospheric angle $\theta_{23}$ differs from $\theta_0$ by a small
angle that can be treated perturbatively as we do in the next section.
In the rotated basis,
\beq
L = \left( \matrix{ 
x_1 & x_- & x_+ \cr 
y_1& y_-& y_+\cr 
z_1& 0 & z_+} \right)\ ,
\eeq
with $z_+ \simeq \sqrt{m_@},$ and all other entries are smaller than
$\sqrt{m_@}$: $x_{\pm}, y_{\pm} \ll \sqrt{m_@}$ expresses the dominance
condition, while $x_1, y_1, z_1 \ll \sqrt{m_@}$ reflects the smallness of
$\tan \theta_{13}.$ How small they should be depends on the
favoured solar neutrino solution and on the value of $\tan \theta_{13}.$

This pattern which is quite suggestive of the properties a), b), c) is
not obtained in straightforward applications of abelian flavour
symmetries to the see-saw mechanism \cite{FN}. But it can be implemented
through the use of holomorphic zeros or non-abelian flavour symmetries
\cite{afm1, bizarre, u2}.

\subsubsection{Inverted hierarchy pattern:
$m_1, m_2 \simeq m_@ \gg m_3$} \label{sec:IHATM}

Since $m^2_2 -m^2_1 =m^2_\odot \ll m^2_@$, $m_1 \simeq \pm m_2$, but
only $m_1 \simeq -m_2$ is consistent with the stability of the
large mixing angle $\theta_{12}$ under radiative corrections. We shall
therefore focus on the case where the mass eigenstates $\nu_1$ and $\nu_2$
form an approximate pseudo-Dirac pair. In this case the structure of the
light neutrino mass matrix indicated by the data -- leaving aside the now
disfavoured small angle MSW solution -- is the following:
\bea
{\cal{M}_\nu}_{11} \sim m_@ ~\max \left( \frac{1-\tan ^2 \theta_{12}}{2},
\frac{m_3}{m_@} U^2_{e3}, \frac{m^2_\odot}{m^2_@} \right)\ , \nn\\
{\cal{M}_\nu}_{22,~33,~23} \sim m_@ ~\max \left( 
\frac{1-\tan^2 \theta_{12}}{2},~U_{e3},~\frac{m_3}{m_@}, \frac{m^2_\odot}{m^2_@}
\right)\ ,  \nn \\
{\cal{M}_\nu}_{13} \sim {\cal{M}_\nu}_{12} \sim m_@\ ,~~~~~~~~~
\frac{{\cal{M}_\nu}_{13}}{{\cal{M}_\nu}_{12}} \simeq - \tan \theta_{23}\ .
\label{ful}
\eea   
Thus ${\cal{M}_\nu}_{12}$ and  ${\cal{M}_\nu}_{13}$ essentially
determine the atmospheric neutrino parameters. The other matrix
elements are related to the solar neutrino parameters and are much
smaller.

Let us identify the patterns of $L$ that are compatible with those
requirements. The more natural way to implement the mass hierarchy between 
the pseudo-Dirac pair and $\nu _3$ is to associate the first  to two 
right handed neutrinos (in the sense that $m_@$
it is dominated by the $z_i$ and $y_i$) while $m_3$ is dominated by the 
$x_i$. Neglecting the contribution of the $x_i$ to
the entries of ${\cal M}_\nu,$ Eqs. (\ref{ful}) can be rewritten as:
\bea
z_1 z_3 + y_1 y_3 & \simeq & m_@ \sin \theta_{23}\ , \nn\\
z_1 z_2 + y_1 y_2 & \simeq & - m_@ \cos \theta_{23}\ , \nn\\
|z_i z_j + y_i y_j| & \ll & m_@  ~~~~~\mbox{for $i=j=1,$  or $i,j\ne 1$}\ .  
\label{sat}
\eea
and also $|x_i x_j| \ll m_@ ~.$ This is precisely the pattern obtained 
in Section (\ref{sec:mechanisms}) where it is shown that the simplest 
realization of a pseudo-Dirac mass structure (\ref{sat}) is when $M_y$ 
and $M_z$ form a pseudo-Dirac pair with some hierarchy in the Yukawa couplings.

If CP is an exact symmetry, we can choose the lepton phases such
that $\cal{M}_\nu $ is a real matrix so that each line in $L$ must have
matrix elements that are all real or all purely imaginary. Therefore
the $z_i$ ($i=1,2,3$) can be taken to be real and the $y_i$ ($i=1,2,3$) 
to be imaginary in order to generate the opposite signs of the masses 
$m_1$ and $m_2 ~,$ also in correspondence with the opposite CP
parities of $M_y$ and $M_z$ (the effects of CP violating phases will be
discussed in more detail later). 

In this inverse hierarchy scenario, it is useful to change to the new
variables: 
\beq
u_i= \frac{z_i + iy_i}{\sqrt{2}} \qquad
v_i= \frac{z_i - iy_i}{\sqrt{2}} \label{uvdef}
\eeq
so that ${\cal{M}_\nu}_{ij} = u_iv_j + u_jv_i + x_ix_j ~.$ The condition
(\ref{psDiY}) reads:
\beq
v_1 \gg v_{2,3} \qquad u_{2,3} \gg u_1 \ .\label{psDiuv}
\eeq
As for the hierarchical spectrum one defines $\pm$ components by the
rotation 
\beq
\left( x_2, ~x_3 \right) \longrightarrow
\left( x_+, ~x_- \right) = 
\left( \cos \theta_0 x_2 - \sin \theta_0 x_3,
~\cos \theta_0 x_3 + \sin \theta_0 x_2 \right)\ ,
\eeq
with $\theta_0 = \arctan (-u_3 /u_2 )$.
After this rotation one has $u_-=0$ and $v_1 u_+ \simeq {m_@}$, which
implies
\beq
{\cal M}_{\nu +-}, ~~ {\cal M}_{\nu 1 -} \ll  
{\cal M}_{\nu 1+} \simeq v_1 u_+ \simeq m_@ ~. \label{psDiconst}
\eeq
The atmospheric angle $\theta_{23}$ differs from $\theta_0$ by a small
angle that can be treated perturbatively as we do in the next section.
In the rotated basis,
\beq
L = \left( \matrix{ 
x_1 & x_+ & x_- \cr 
u_1& u_+ & 0 \cr 
v_1& v_+ & v_-} \right)\ , \label{Luv}
\eeq
The symmetry of the solutions under $y_i \rightarrow -y_i$ corresponds
here to $u_i \leftrightarrow v_i ~.$ 

There are three possibilities consistent with (\ref{psDiconst}): 1)
$|v_1| > \sqrt{m_@} > |u_+| \ ,$ 2) $|u_+| > \sqrt{m_@} >|v_1|\ ,$ 3)
$|u_+| \sim |v_1|\sim \sqrt{m_@}.$ Since these three kinds of inverted
hierarchy solutions correspond to the same pattern for  $\cal{M}_\nu$,
they should be connected by orthogonal matrices $R$. If we use the
parameterization (\ref{laLbis}) with ${\tilde R}=1$, one obtains from the
conditions (\ref{sat}) the solutions of type 2).  A {\it boost}, \ie~  
a rotation by an imaginary angle $i\lambda$ in the $(y,z)$ sector
transforms the variables $(u,v)$ as
\beq
u_i \rightarrow e^{\lambda} ~u_i \qquad 
v_i \rightarrow e^{-\lambda} ~v_i \qquad u_iv_j \rightarrow u_iv_j \ .
\label{boost}
\eeq
When $\lambda$ is large enough, a solution of type 1) is transformed
into a solution of type 2). In between, the solutions 3) can be
obtained. These orthogonal transformations (\ref{boost}) parameterize a
continuum of equivalent mathematical solutions with quite different
physics contents (see Section \ref{sec:CLFV}). Notice that in each case
$M$ is defined to be diagonal and does not commute with $R$.

The elements of the matrix $L$ are naturally bounded by ${\cal O} (v/
M_i^{1/2})$, as far as the Yukawa couplings are perturbative at the
scale of lepton number violation.  Therefore, if e.g. some $y_i \sim
\sqrt{m_@},$ one has $M_y \lesssim v^2/m_@ \sim 5 \times 10^{14} $ GeV,
and analogously for the $x_i$ and $z_i$. In the hierarchical case, one
has (with our convention that $m_@$ is dominated by $M_z$) $ M_z
\lesssim  5 \times 10^{14} $ GeV, while $M_x$ and $M_y$ remain free.
In the inverted hierarchy case, there are two possibilities: for the
case 3), $M_y$ and $M_z$ are bounded by $5 \times 10^{14}$ GeV, while
in cases 1) and 2) the upper bound is lower since either $v_1$ or $u_+$
is larger than $\sqrt{m_@}$.

Now that the large scale ${m_@}$ has been identified, the oscillation
parameters $\theta_{13},$ $\theta_{12},$ $\theta_{23}-\theta_{0}$ and
$m_\odot$ can be determined by an expansion in these small parameters
that are constrained by experimental data in turn. This is discussed in
the next Section, where CP violating phases are also taken into
account.

\section{Origin of The Large Solar Angle}\label{sec:solar}

As discussed in the previous Section, only two patterns for
$\cal{M}_\nu$ are consistent with the requirement of a large
atmospheric mixing and $m^2_@ >> m^2_{\odot}$. Along this
line of reasoning, the next step is to investigate whether the
requirement of a large - or possibly maximal - solar angle can give
further informations on the sub-leading terms of those two patterns.  It
is important to take into account the CP violating phases potentially
present in $Y$ and so in $L ~.$ In the following we carry out this
program in a quantitative and systematic way. Of course some results
are already known as empirical rules from many previous analyses.
Fortunately, a more quantitative reappraisal is already possible with
reliable approximate expressions.

In this section we adopt the following notation:
\beq
A \approx B \ \ \Longleftrightarrow \ \ A = B
~\left( 1+ {\cal O}\left(U_{e3}^2,{m_{\odot}}/{m_@}\right) )\right)\ ,
\eeq
which relates the neglected contributions to the physically small
parameters.

We are seeking for sufficiently well approximated expressions for the
observables contained in $\cal{M}_\nu$, in particular for the three mixing
angles and the CP violating phases of the MNS matrix $U,$
which can be written in the form
\beq
U=e^{i \alpha}~{\cal W}~R(\theta_{23}) R(\theta_{13})\,
\mbox{diag} (1, e^{-i \chi}, 1) R(\theta_{12}) ~{\cal V}  ~. 
\eeq 
where $\cal W$ and $\cal V$ are diagonal matrices of $SU(3)$. However,
in the interaction basis where charged lepton masses are diagonal, there
is still the freedom of phase redefinition for each one of the leptons
$e,\mu,\tau$; at the level of the matrix $L$ this amounts to the freedom of
multiplying each column by an arbitrary phase.

Therefore $\alpha$ and the phases in $\cal W$ are not physical as they can
be absorbed in the $e,\mu,\tau$ (on the contrary, the two Majorana phases
in ${\cal V}$ and the CKM-like phase $\chi,$ often denoted by $\delta$ in
the literature, are physical; in particular $\chi$ could be measured in
oscillations at a neutrino factory).
Once this is done, it is convenient to define the matrix
\beq
\ell \equiv L R(\theta_{23}) R(\theta_{13})~.
\eeq
Indeed the effective light neutrino mass matrix reduces, in the basis defined
by $\ell,$ to
\beq
m \equiv \ell^T \ell = \left( \matrix{ m_{11} & m_{12} & 0 \cr 
m_{12} & m_{22} & 0 \cr 0 & 0 & m_3 } \right)\ , \label{defm}
\eeq
and the solar neutrino parameters $ m^2_{\odot}$ and $\theta_{12},$
as well as the CP-violating phases $\chi$ and ${\cal V},$ can be obtained by
diagonalizing a $2 \times 2$ matrix.

The CKM-like phase $\chi$ is determined by the condition
\beq
| {m}_{22} | \sin (\alpha_{2} -\chi)\ =\
| {m}_{11} | \sin (\alpha_{1} +\chi)\ , \label{sinus}
\eeq
and the solar angle is given by
\beq
\tan 2 \theta_{12}\ =\ \frac{2|{m}_{12}|}
{|{m}_{22}| \cos (\alpha_{2}  -\chi)-
 |{m}_{11}| \cos (\alpha_{1} +\chi) }\ ,  \label{theta12}
\eeq
where $\alpha_{2}=\arg(m_{22})-\arg(m_{12})$ and $\alpha_{1} =
\arg(m_{11})-\arg(m_{12})$. Eq. (\ref{sinus}) has two solutions
which differ by $\pi,$ and correspondingly $\tan 2 \theta_{12}$ can
take both signs. However the two solutions are physically equivalent,
as expected, since for $0 \leq \theta_{12} \leq \pi/2$ they are
related by exchanging $m_1$ and $m_2.$ Thus, it is not restrictive to take 
$m^2_\odot \equiv \Delta m^2_{21} \equiv m^2_2 - m^2_1 > 0,$ so that $|m_{22}| > |m_{11}|$ 
corresponds to $0 \leq \theta_{12} < \pi/4$ while $|m_{22}| < |m_{11}|$
corresponds to $\pi/4 < \theta_{12} \leq \pi/2$ (``dark side'').

Eq. (\ref{sinus}) is graphically solved for $\chi$ by the construction
of the two triangles in Fig. 1. The other phase dependent quantities
discussed in this section, in particular the denominator of
(\ref{theta12}), are also shown in the figure. The CP-conserving case
corresponds to a flat triangle.  One can easily see that, for fixed
$|m_{11}|$ and $|m_{22}|,$ the maximal (resp. minimal) cancelation in
the denominator of Eq. (\ref{theta12}) is obtained in the CP-conserving
case with $m_{11} m_{22} > 0$ (resp. $m_{11} m_{22} < 0$). Thus the
general effect of the CP-violating phases is to reduce the solar mixing
angle with respect to the CP-conserving case with $m_{11} m_{22} > 0$.

As for $m^2_{\odot},$ one has one phase independent relation,
\beq
m^2_{\odot}\cos 2\theta_{12} = |m_{22}|^2 - |m_{11}|^2 \label{msol}
\eeq
and another one is  
\beq
  m^2_{\odot} \sin 2 \theta_{12}\ =\ 2 |m_{12}| \left( |{m}_{22}|
  \cos (\alpha_{2} - \chi) + |{m}_{11}| \cos (\alpha_{1} +\chi) \right)\ ,
\label{msol_bis}
\eeq
which basically gives $m^2_{\odot}$ for near maximal mixing.  

\begin{figure}[!t]
\centerline{\psfig{file=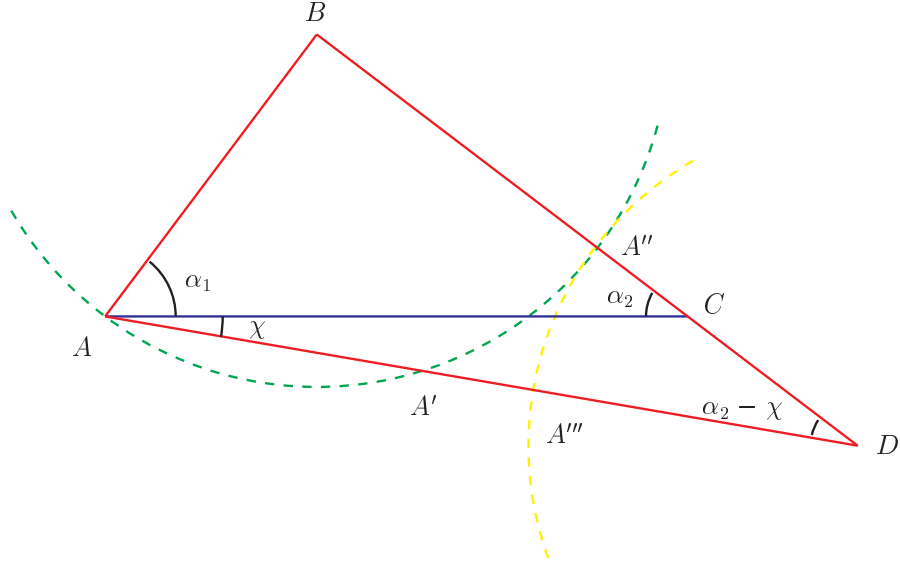,width=0.8\textwidth}}
\caption{The triangle defined by the condition on $\chi$ eq. (\ref{sinus}):
$AB=A'B=A''B=|m_{11}|$, $BD=|m_{22}|$. The denominator of eq. (\ref{theta12}) is
$A'D=|m_{22}| \cos(\alpha_2 -\chi)-|{m}_{11}| \cos (\alpha_{1} +\chi) > A''D$, where 
$A''D=A'''D=|m_{22}|-|m_{11}|$. Thus, phases work against the large mixing.}
\label{triangle}
\end{figure}

Two effects can conspire to give a large mixing angle:  $|m_{12}|
\gtrsim |m_{11}|, |m_{22}|$ and some amount of cancelation in the
denominator of Eq. (\ref{theta12}), with $|{m}_{22}| \cos(\alpha_{2} -
\chi) \sim |{m}_{11}| \cos(\alpha_{1} + \chi).$ If we combine
(\ref{theta12}) and (\ref{msol_bis}) into
\beq 
\frac{m^2_{\odot} \tan^2 \theta_{12}}{|m_{12}|^2 
| 1 - \tan^4 \theta_{12} |} =\left|  
\frac{|{m}_{22}| \cos (\alpha_2 -\chi)+|{m}_{11}|\cos (\alpha_1 +\chi)}
{|{m}_{22}| \cos (\alpha_{2} -\chi)-|{m}_{11}| \cos (\alpha_{1} +\chi)} \right| \gtrsim 1 \ , 
\label{test}
\eeq
one has a criterion to decide which mechanism is preferred. Indeed, if
the denominator in (\ref{theta12}) is made small by cancelation, the
ratios in (\ref{test}) are expected to be much large. Instead, if both
terms in this denominator are small, these ratios are ${\cal O}(1).$
This criterion will be applied to different possible seesaw patterns
here below, by first obtaining also constraints on $|m_{12}|.$

\subsection{Hierarchical Spectrum}

Let us first consider the case of hierarchical neutrinos. As already
discussed, the associated pattern of $\cal{M}_\nu$ suggests that there
are two dominant elements in $L$, which we are free to place in the
third row of $L$ -- namely $z_{2,3}$. The freedom linked to the phases
$\cal W$  can be used to ensure that $\theta_{23}$ and $\theta_{13}$
are real. The expressions for these angles can be written as 
\beq
\tan{\theta_{23}}  \approx  \frac{z_2}{z_3}
                    +  \frac{y_- \bar{y}_+ + x_- \bar{x}_+}{z_3^2}\ ,
\label{t23}
\eeq
\beq
\tan{\theta_{13}} \approx \frac{z_1}{z_+} 
 + \frac{\bar{y}_1 \bar{y}_+ + \bar{x}_1 \bar{x}_+}{z_+^2}\ ,  \label{t13}  
\eeq
where the variables $(x_{\pm},~y_{\pm},~z_{\pm})$ are defined in
(\ref{rotaz}), $\bar{y}_+ =y_+ +(z_1/z_+) y_1$ and $\bar{y}_1 =y_1-
(z_1/z_+) y_+$, and analogously for $\bar{x}_+\, ,~\bar{x}_1$.
Therefore the two phases in $\cal W$ are chosen to eliminate the
overall phases in these two expressions, which are independent of the
phase $e^{i\alpha} ~.$

After performing these rotations in this approximation, one gets 
\bea
\ell = L R(\theta_{23}) R(\theta_{13}) \approx  \left( 
\matrix{ \bar{x}_1 & x_- & \bar{x}_+ \cr
\bar{y}_1 & y_- & \bar{y}_+ \cr  
-\frac{\bar{x}_1 \bar{x}_+ +\bar{y}_1 \bar{y}_+}{z_+} & 
-\frac{ x_- \bar{x}_+ + y_- \bar{y}_+}{z_+}  & z }\right)\ , 
\eea
where $z^2 \equiv \sum_i z_i^2 \approx m_@ $. Then $m$ reads
\bea
m \approx \left( \matrix{
\bar{y}_1^2+\bar{x}_1^2 & \bar{y}_1 y_- +\bar{x}_1 x_- &  0 \cr
\bar{y}_1 y_- +\bar{x}_1 x_-  & y_-^2 + x_-^2  &  0 \cr
0  &  0 & z^2+\bar{y}_+^2+\bar{x}_+^2 } \right)\ .  \label{m_hierarchy}
\eea

By replacing these matrix elements in (\ref{sinus}),(\ref{theta12}),
(\ref{msol}) and (\ref{msol_bis}), one obtains the expressions for
${\tan}^2 \theta _{12}$ and $m_{\odot}.$ In the following, we shall
describe separately the two mechanisms that can lead to a large
atmospheric mixing angle: cancelation in the denominator of Eq.
(\ref{theta12}) and $|m_{12}| \gg |m_{11}|, |m_{22}|$).  One can see
from Eq. (\ref{m_hierarchy}) that the latter can be obtained only at
the price of a cancelation between $\bar x^2_1$ and $\bar y^2_1$,
and/or $x^2_-$ and $y^2_-$; in this case $\nu_1$ and $\nu_2$ form a
pseudo-Dirac pair.

As far as $|m_{12}| = \bar{x}_1 \bar{x}_- +\bar{y}_1 \bar{y}_- $ is
concerned, one can use the fact that, generically, \ie\ barring a close
correlation between the $z$'s and $y$'s, one has $|y_+| \sim |y_-| \sim
\max\, (|y_2|, |y_3|),$ and analogous relations for the $x$ components.
Now from (\ref{t13}) one gets the bound
\beq
|m_{12}| \sim \left|\bar{x}_1 \bar{x}_+ +\bar{y}_1 \bar{y}_+ \right|
\lesssim |U_{e3}| m_@ \label{m12bound}
\eeq
that can be used in (\ref{test}) to test the large mixing
mechanism in hierarchical models as discussed below in more
detail.

If $m_@$ is dominated by one right-handed neutrino and the solar mixing
angle is due to some cancelation between the two terms in the
denominator of Eq.  (\ref{theta12}), there are two situations as far as
the two remaining neutrino mass eigenvalues are concerned. They can
exhibit some hierarchy, with $m_2 \simeq m_{\odot} \gg m_1$, or they
can be comparable in magnitude (but not much larger  than $m_@$ to keep
large mixing angles stable under the RGE evolution). Let us discuss the
first possibility in detail and then comment on the second one.

\subsubsection{Double Dominance}\label{sec:dd}

In the presence of the large solar angle, the most natural realization
of a hierarchy between $m_2$ and $m_1$ involves a dominance mechanism
analogous to the one assumed in the atmospheric sector. Indeed,
if $\bar x_1 \ll \bar y_1,$ $x_- \ll y_-$, the solar neutrino scale is
dominated by the right-handed neutrino with mass $M_y,$ namely
$m_{\odot} \simeq m_2 \approx |y_1^2 + y_-^2|$ (the choice of $x$ negligible with respect to $y$
is not restrictive: $\bar y_1 \ll \bar x_1,$ $y_- \ll x_-$
corresponds just to the exchange $M_x \leftrightarrow M_y$).
When $\bar x_1, x_-$ are much smaller than $\bar y_1, y_-$ the formulae above simplify considerably
also because sines and cosines factorize out. Indeed, it turns out from the condition (\ref{sinus}) that 
$\alpha_2-\chi \approx -(\alpha_1+\chi)=0$. The CP violating phase $\chi$ is 
\beq
\chi\ \approx\ \arg(y_-) - \arg(\bar y_1). 
\eeq
The expressions for the solar angle and the solar mass scale become simply
\beq
  \tan \theta_{12}\ \approx\ r\ , \qquad
  m_{\odot}\ \approx\ |y_-|^2 \left( 1+r^2 \right)\ , \qquad
  r\ \equiv\ \frac{|\bar y_1|}{|y_-|}\ . \label{ddrelations}
\eeq
Then, as was the case for the atmospheric angle, a large solar angle results if the ratio of
two couplings, $\bar y_1/y_-$, is of order one. So, the prediction
for a large solar angle is stable in the sense specified in the introduction. 
Together with the assumed dominance in $m_@,$ this corresponds to a double-dominance pattern (or
triple-dominance pattern, since the lightest mass eigenvalue, $m_1,$ is
then dominated by the third right-handed neutrino).
LMA requires $0.5 < r < 0.9$ and $|y_-|^2 = (3-8) \times 10^{-3}$ eV,
while LOW requires $0.7 < r < 1.1$ and $|y_-|^2 = (1-3) \times 10^{-4}$
eV (using the 99\% C.L. intervals of Ref. \cite{bach}).  A
maximal solar angle would require a tuning of $r$ to $1.$ Finally the
Majorana phase in ${\cal V}$ associated with $m_2$ is $\lambda_2 \equiv
\arg({\cal V}_{22}) \approx \arg(\bar y_1).$ In the limit $m_1
\rightarrow 0,$ there is no other CP-violating phase; in general
however $U$ contains a third CP-violating phase associated with $m_1$,
$\lambda_1 \equiv \arg({\cal V}_{11})$ (in our phase conventions,
$\arg({\cal V}_{33}) = 0$).

\begin{figure}[!ht]
\center{a) Double dominance~~~~~~~~~~~~~~~~~~~~~~~~~~~~~~~~~~~~~~~~~~~~b) Pseudo Dirac}
\centerline{\psfig{file=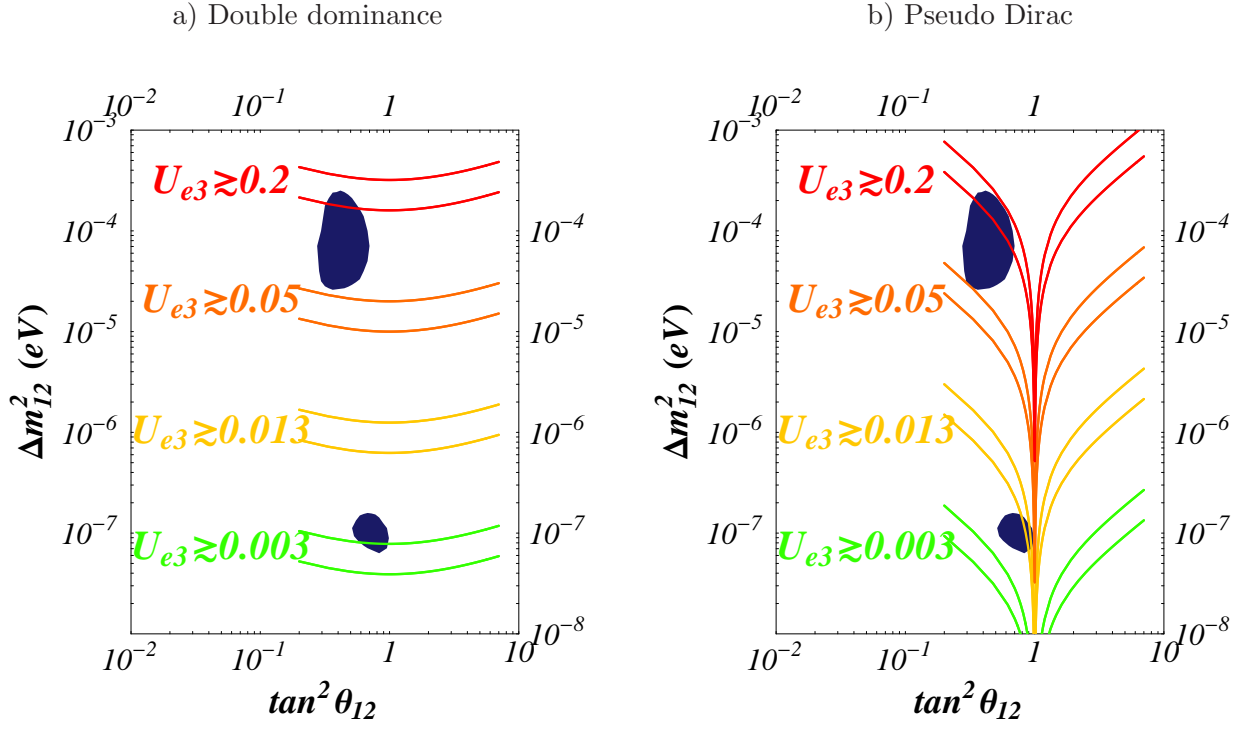,width=1\textwidth}}
\caption{Lower limits on $U_{e3}$ for double dominance a) and pseudo-Dirac b). To sketch the regions of 
the LMA and LOW solutions we have shown in blue the 95\% C.L. contours of the first reference
of \cite{fit_including_SNO}.
The lower (upper) curve displayed for each color corresponds to $m_@^2= 2~ (4) \times 10^{-3}$ eV$^2$.}
\label{Hiue3}
\end{figure}

A very constraining conditions holds for the double dominance pattern
which follows from the (\ref{test}), (\ref{m12bound}) and
(\ref{ddrelations})
\beq
  |U_{e3}|\ \gtrsim\ \frac{1}{2} \sin 2 \theta_{12}\, \frac{m_{\odot}}{m_@}\ .
\label{ddlim_LA}
\eeq
For LMA this means that $U_{e3}$ should be within one order of magnitude
from its present experimental limit, while it can be much smaller for LOW,
as shown in Figure \ref{Hiue3} a).

We finally note, from the assumed perturbativity property of the Yukawa
couplings, that $M_z \lesssim 5 \times 10^{14}$ GeV, while $M_y
\lesssim \frac{m_@}{m_{\odot}} \times 5\times 10^{14}$ GeV. For the LOW
solution $M_y$ can lie above the unification scale (when the
corresponding Yukawa couplings are of order one).

The double dominance pattern can also accommodate a small solar angle,
with $\bar y_1 \ll y_-$ instead of $\bar y_1 \approx y_-.$ For the (now
strongly disfavoured by the data) SMA solution, one needs $0.01 < r <
0.03$ and $|y_-|^2 = (2-3) \times 10^{-3}$ eV. The CHOOZ angle is
constrained to be small; more precisely, it lies in the following
range:
\beq
  \tan \theta_{12}\, \frac{m_{\odot}}{m_@}\ \lesssim\ |U_{e3}|\
  \lesssim\ \tan \theta_{12}\ .
\label{ddlim_SMA}
\eeq
Thus $|U_{e3}|$ is at least one order of magnitude smaller than its present
experimental limit; this is to be contrasted with the large angle solutions
for which there is no theoretical upper bound on $|U_{e3}|$.

\subsubsection{Pseudo-Dirac Solar Neutrinos}

The alternative option for generating a large solar angle is to assume a
pseudo-Dirac structure with $|m_{11}|, |m_{22}| \ll |m_{12}|,$ \ie
\beq
  |\bar y^2_1 + \bar x^2_1|\, ,\, |y^2_- + x^2_-|\ \ll\
  |\bar y_1 y_- + \bar x_1 x_-|\ \equiv\ \bar m\ .
\label{eq:PD_hierarchy}
\eeq
This pattern leads to a strongly mixed pseudo-Dirac pair of neutrinos
with masses $m_2 \simeq - m_1 \simeq \bar m;$ $ |U_{e3}| m_@ \gtrsim
\bar m \gg m_{\odot}.$ As already discussed in Section
\ref{sec:seesaw}, the conditions (\ref{eq:PD_hierarchy}) require
cancelations between the $y$'s and the $x$'s, which is most easily
implemented with a pair of pseudo-Dirac right-handed neutrinos, with
$M_y \simeq - M_x \simeq \bar{M},$ together with some hierarchy among
the Yukawa couplings.

Since a large angle is an automatic consequence of the smallness of
$|m_{11}|,$ $|m_{22}|$ implied by (\ref{eq:PD_hierarchy}),
no particular correlation between them is required, and the ratios in the
test relation (\ref{test}) are ${\cal O}(1).$ Replacing $|m_{12}|$ by its
upper bound $|U_{e3}| m_@$ one finds the constraint
\beq
  \frac{m^2_{\odot}} {m^2_@}  \frac{\tan^2 \theta_{12}}  { |1 - \tan^4 \theta_{12}|}\
  \lesssim |U_{e3}|^2\ .
\label{eq:CH00Z_PD}
\eeq
As shown in fig. \ref{Hiue3} b), in the LMA region 
$|U_{e3}|$ should be very close to its experimental bound and saturate the lower
bound in Eq. (\ref{eq:CH00Z_PD}). There is more room for the LOW solution. 
The lower bound on $|U_{e3}|$ in Eq. (\ref{eq:CH00Z_PD}) depends on how
$\theta_{12}$ is close to $\pi/4;$ the smaller $1 - \tan^2 \theta_{12},$ the larger $|U_{e3}|.$

From the perturbativity property of the Yukawa couplings one can derive
an upper bound on the right-handed pseudo-Dirac neutrino mass, $\bar{M}
\lesssim \frac{m_@}{\bar{m}} \times 5 \times 10^{14}$ GeV.  Since
$\frac{m_@}{\bar{m}} \gtrsim |U_{e3}|^{-1},$ the pseudo-Dirac
right-handed neutrino pair could be heavier than the third right-handed
neutrino, whose mass satisfies $M_z \lesssim 5 \times 10^{14}$ GeV.

\subsubsection{Two-scale spectrum: $m_1 \sim m_2$}

The double dominance mechanism gives a fully hierarchical mass
spectrum, $m_1 \ll m_2 \ll m_3.$ However, $m_1 \sim m_2$ is in principle a viable
option too. This corresponds to the case where no single right-handed
neutrino dominates in the left upper $2 \times 2$ submatrix in Eq.
(\ref{m_hierarchy}). Then, full Eqs. (\ref{sinus}) to
(\ref{msol}) must be used, and there is no simple correspondence
between the solar neutrino parameters and the couplings of one right-handed
neutrino as in the case of double dominance. The large solar angle requires
$|{m}_{22}| \cos(\alpha_{2} - \chi) \sim |{m}_{11}| \cos(\alpha_{1} + \chi)$
and $|m_{12}| \sim |m_{11}|, |m_{22}|,$ (which is obtained for
(i) $\bar x_1 \lesssim \bar y_1 \sim y_-$ if $x_- \ll y_-$;
(ii) $\bar y_1 \lesssim \bar x_1 \sim x_-$ if $y_- \ll x_-$;
(iii) $\bar x_1 \lesssim \bar y_1 \sim y_-$ or
$\bar y_1 \lesssim \bar x_1 \sim x_-$ if $x_- \sim y_-$), and 
the solar neutrino scale is approximately given by
$m_{\odot} \sim |y^2_- + x^2_-| \cos(\alpha_{2} - \chi)
+ |\bar y^2_1 + \bar x^2_1| \cos(\alpha_{1} + \chi)$.
Thus, sensible variations in the magnitude of $\theta_{12}$ and $m_{\odot}$ are
expected by slightly perturbing the many parameters entering the previous
expressions, in particular the phases. Indeed, as already mentioned, the phases
work against a large solar angle (they are not dangerous in the double dominance case 
because the cosinus in the denominator of eq. (\ref{theta12}) factorize out).   
Thus, the two-scale spectrum cannot arise from the see-saw in an economical and robust 
way. 
The CHOOZ angle satisfies again 
the same lower bound as in the double dominance case:
$  |U_{e3}|\ \gtrsim\ {\cal O} \left( m_{\odot}/m_@ \right).$
However, due to the large number of parameters involved, this bound is less
strict than Eq. (\ref{ddlim_LA}). 
For completeness, note that the SMA solution would be accommodated with
(i) $\bar y_1 \ll y_-$ and $\bar x_1 < y_-$ if $x_- \ll y_-$;
(ii) $\bar x_1 \ll x_-$ and $\bar y_1 < x_-$ if $y_- \ll x_-$;
(iii) $\bar x_1 \ll x_-$ and $\bar y_1 \ll y_-$ if $x_- \sim y_-$.
The solar neutrino scale is then given by $m^2_{\odot} \simeq
|y^2_- + x^2_-|^2 - |\bar y^2_1 + \bar x^2_1|^2,$ and the CHOOZ angle
lies in the same range as in the double dominance case if both
$\bar x_1, \bar y_1 \ll \max (x_-,y_-).$

\subsection{Inverted Hierarchical Spectrum} \label{sec:iHsolar}

Let us now discuss the alternative case of an inverse hierarchy in the
neutrino mass spectrum. We shall work with the variables $(x_i ~, ~u_i
~, ~v_i),$  introduced in section (\ref{sec:patterns}) such that $L$
is given by (\ref{Luv}). Up to irrelevant phase redefinitions, the
implementation of the atmospheric neutrino oscillations require $v_1
u_+ \simeq {m_@} ~,$ with the other components much smaller. As before,
the unphysical phases $\cal W$  can be chosen so that $\theta_{23}$ and
$\theta_{13}$ are real. Expanding in the smaller matrix elements of $L$,
one obtains the following expressions for these angles:
\beq
\tan{\theta_{23}} \approx -\frac{u_3}{u_2} - \left( \frac{u_+^2}{u_2^2}  
\right) \frac{\bar{v}_1 \bar{u}_- +\bar{x}_1 \bar{x}_- }{\bar{v_1}u_+} 
\approx  -\frac{u_3}{u_2} - \left( \frac{u_+^2}{u_2^2}  
\right) \frac{- v_-^2+ x_1 x_- }{m_@}  ~,  \label{psDit23}
\eeq
\beq
\tan{\theta_{13}} \approx -\frac{v_-}{v_1} + 
\left(\frac{\bar{v_1}^2}{v_1^2}\right)
\frac{\bar{u_-} v_+ + \bar{x_-} x_+}{\bar{v_1}u_+}
\approx  -\frac{v_-}{v_1} \left( 1 + \frac{u_1v_+}{m_@} \right) 
+ \frac{x_- x_+}{m_@}  
\label{psDit13}  
\eeq
where the variables $(x_{\pm},~v_{\pm},~u_{+})$ are defined in
(\ref{rotaz}), with $\theta_0 = \arctan (-u_3 /u_2 )$, $u_- = 0$, and
\bea 
\bar{x}_1 {\bar{v}_1} &=& (v_1 x_1 + v_- x_- ) ~,~~~~ 
\bar{x}_- {\bar{v}_1} ~=~ (v_1 x_- - v_- x_1 ) ~,~~~~ 
\bar{u}_1 {\bar{v}_1} ~=~ v_1 u_1 ~,  \cr
\bar{u}_- {\bar{v}_1} &=& - v_- u_1 ~, ~~~~~~
\bar{v}_- ~=~ 0 ~,~~~~~~\bar{v}_1^2 ~=~ v_1^2 + v_-^2 ~,~~~~~~
u_+ \bar{v}_1 ~\approx ~m_@ ~. 
\eea
In this approximation, the matrix $m$ defined in (\ref{defm}) reads
\bea
m \approx \left( \matrix{
2u_1 v_1+\bar{x}_1^2 & \bar{v}_1 u_+ +\bar{u}_1 v_+ +\bar{x}_1 x_+ & 0\cr
\bar{v}_1 u_+ +\bar{u}_1 v_+ +\bar{x}_1 x_+ & 2u_+ v_+ +x_+^2 &  0 \cr
0  &  0 & \bar{x}_-^2 } \right)\ .  \label{m_iH}
\eea
By replacing these matrix elements in (\ref{sinus}),(\ref{theta12}),
(\ref{msol}) and (\ref{msol_bis}), one obtains the expressions for
${\tan}^2 \theta _{12}$ and $m_{\odot}.$

Since the matrix $m$ and the oscillation parameters discussed in this
section are invariant under the boost given in (\ref{boost}), it is
worth defining  the {\it rapidity} $\lambda = 
\frac{1}{2} \log (u_+ /v_1 )$, such that
\beq
u_+ \simeq e^{\lambda} \sqrt{m_@} \qquad 
v_1 \simeq e^{-\lambda} \sqrt{m_@}  \ .\label{lambda}
\eeq
while the parameters $(u_1 /u_+ ),\, (v_+ /v_1 ),\, (v_- /v_1 ) ,\,$
are $\lambda$-independent. They appear in the approximate expressions:
\bea 
\frac{m^2_{\odot}}{2m_@^2} &\simeq & 2\frac{u_1}{u_+} +2\frac{v_+}{v_1}
+\frac{x_1^2 + x_+^2}{m_@}  \cr
\tan ^2{\theta _{12}} - 1 &\simeq & 2\frac{u_1}{u_+} -2\frac{v_+}{v_1}
+\frac{x_1^2 - x_+^2}{m_@}  \ ,\label{psDiapprox}
\eea
where the phase dependence discussed in Section \ref{sec:solar} has 
been omitted for simplicity.
 
\begin{figure}[!ht]
\centerline{\psfig{file=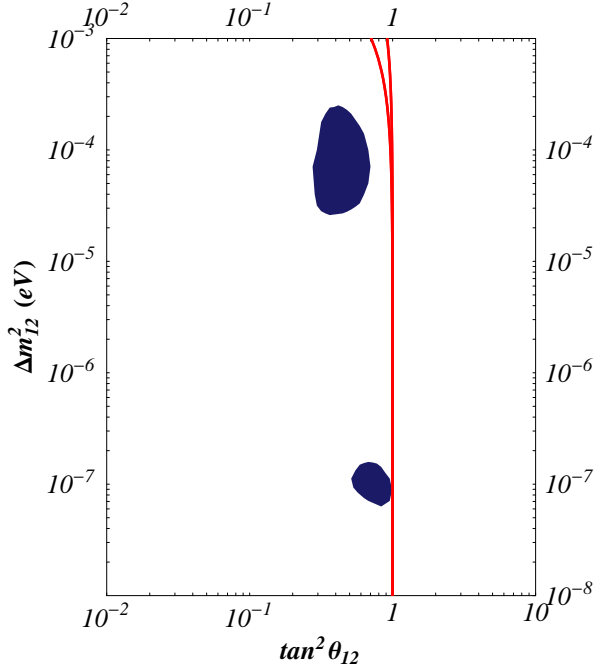,width=0.5\textwidth}}
\caption{The curve determined by Eq. (\ref{iHtest}). The blue regions are like those of Fig. \ref{Hiue3}.}
\label{iH}
\end{figure}

The important feature here is the relation $|m_{12}| \approx v_1 u_+
\approx m_@ $. Since in this inverse hierarchy scenario, the large
solar neutrino mixing is related to the pseudo-Dirac structure, we
preclude a strong cancelation in the right-hand side of the relation
(\ref{test}), namely a precise cancelation between the matrix elements
$m_{11}$ and  $m_{22}$ of (\ref{m_iH}). In particular, the two
expressions in (\ref{psDiapprox}) must be of the same order.  Thus, we
derive from (\ref{test}), by replacing $|m_{12}|$ by $m_@ ,$ the
constraint
\beq 
\frac{m^2_{\odot}}{m_@^2} \sim 
\frac{\left( 1-\tan ^4 \theta_{12}\right)}{\tan ^2\theta_{12}} 
\ , \label{iHtest}
\eeq
The comparison with the fits to neutrino experiments is displayed in 
Fig. \ref{iH}. Therefore the inverse hierarchy patter is strongly disfavoured
for LMA unless we complement the pseudo-Dirac pattern with a fine tuning
of $m_{11}$ and  $m_{22} .$ As for the LOW solution one needs the solar 
mixing angle
to be very close to its maximal value, namely,  $|1-\tan ^2\theta_{12}|
< 10^{-4} .$ 

Generically, \ie\  without fortuitous cancelations, one expects that 
these strong upper limits on $m_{11}$ and  $m_{22}$ will translate into
correspondingly small values for $|v_+ / v_1 |,\ |u_1 / u_+ |$ and
$|x_1^2 |/m_@ ,| x_+^2 |/m_@ \ .$ As already argued, we assume the
natural relations $|v_- | \sim |v_+ |$ and  $|x_- |\sim |x_+ | ,$
to obtain a limit 
\beq
|U_{e3}| \lesssim \frac{m^2_{\odot}}{2m_@^2} \label{psDiUe3} 
\eeq
which is below the present experimental limits for LMA
and well below  it for LOW. Notice that this conclusion is quite 
independent of the tuning to get $m_{\odot}$ discussed above.

Actually, one can distinguish two typical mechanisms to implement the
two small scales $m_{\odot}$ and $m_3 \approx \bar{x}_-^2 ~$ as well as
the small deviation from maximal solar mixing angle $\theta_{12}$:

\noindent 1) {\it Double dominance by a pseudo-Dirac pair,} so that
$m_{\odot}>m_3$: when the contributions of the terms of the form $x_i
x_j$ can be neglected in (\ref{psDit23}), (\ref{psDit13}) and
(\ref{psDiapprox}). Both the atmospheric and the solar neutrino
oscillations are basically controlled by the sector $(u~ , ~v),$ hence
$m_{\odot}$ and $\theta _{12}$ are dominated by the right-handed
pseudo-Dirac pair contributions. In this case, the ratios $(u_1 /u_+
),\, (v_+ /v_1 ),\, (v_- /v_1 ) ,\,$ are approximately determined by
(\ref{psDit13}) and (\ref{psDiapprox}).

\noindent 2) {Single dominance of} $m_{\odot}\, ,$ with $m_{\odot}\sim
m_3$: when the contributions of the terms of the form $u_i v_j$ can be
neglected in (\ref{psDit23}), (\ref{psDit13}) and (\ref{m_iH}) (but for
$u_+ v_1$), so that the terms  $x_i x_j$ control the solar parameters
$m_{\odot}$ and the deviation of $\theta _{12}$ from $\pi / 4\, ,$ with
a dominance by the third right-handed neutrino.  Here, $x_1,\, x_-,\,
x_+\, ,$ are approximately fixed by (\ref{psDit13}) and
(\ref{psDiapprox}).

As we now turn to discuss, charged lepton flavour violating (CLFV) decays can provide some discrimination
between these different possible patterns.

\section{Charged lepton flavour violating radiative decays} \label{sec:CLFV}

In the previous section, we identified the patterns
for $L$ that reproduce in the most natural way, from a bottom-up point
of view, the experimental results. 
Of course, neither the absolute scale nor the hierarchy of right-handed neutrino 
masses can be deduced from oscillation data.
Other experiments are needed to gain further insights; 
in particular, searches for flavour-violating
charged lepton decays such as $\tau \rightarrow \mu \gamma$ and
$\mu \rightarrow e \gamma$ offer the possibility to test another combination
of $Y$ and $M$ than the one associated with oscillations.

It is well known that the flavour structure of soft terms in the slepton
sector can lead to strong violations of lepton flavour in
supersymmetric extensions of the Standard Model.
Even if one assumes that the mechanism of supersymmetry breaking
is flavour-blind, with ${m^2_{\tilde L}}_{ij} = {m^2_{\tilde e}}_{ij}
= m^2_0 \delta_{ij}$ and $A^e_{ij} = A_0 Y^e_{ij}$ at some
scale $M_U \sim M_{Pl},$ the soft terms can receive additional flavour-dependent
contributions from various sources, which may lead to an observable rate
for processes such as $\tau \rightarrow \mu \gamma$ and
$\mu \rightarrow e \gamma.$ Among the possible contributions are the
radiative corrections induced by the
right-handed neutrino couplings $Y_{ki}$ \cite{Borzumati86}, 
radiative corrections in the
context of grand unification\footnote{For instance it is well known that in
$SU(5)$ an additional source of LFV are the Yukawa couplings of the
colored triplet \cite{Barbieri95}.} and other possible contributions
from unknown physics between $M_{GUT}$ and $M_{Pl}$.

\begin{figure}[!ht]
\centerline{\psfig{file=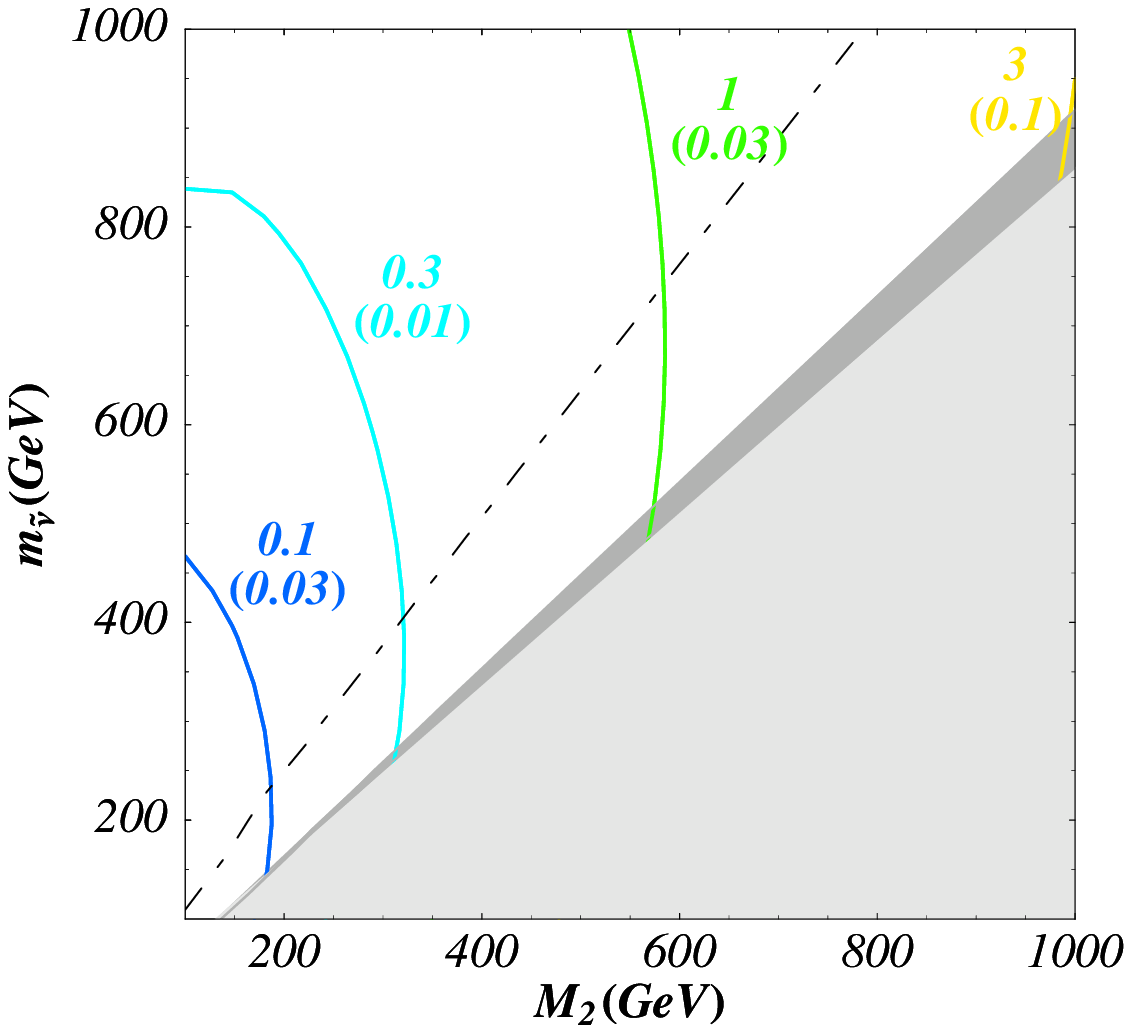,width=0.8\textwidth}}
\caption{Upper bound on $\frac{(m^2_{\tilde \nu})_{\mu \tau}}
{m^2_{\tilde \nu}}$ from the present experimental limit
$BR(\tau \rightarrow \mu \gamma) < 1.1 \times 10^{-6}$ as a function of
the mean sneutrino mass $m_{\tilde \nu}$ and of the $SU(2)_L$ gaugino soft
mass $M_2,$ for $\mbox{tan}\beta=10$, $A_0=0$ and $\mbox{sign} (\mu)=+.$
The numbers in brackets correspond to an improvement by a factor $10^3$
in the experimental limit on $BR(\tau \rightarrow \mu \gamma)$.
The dark gray region is excluded because the lighter stau would here be the
LSP, while the light grey region corresponds to $m^2_0 < 0$.
The region below the dash-dotted line could be explored by searching for LFV
slepton decays at future colliders \cite{Hinchliffe01}.}
\label{figtmg_1}
\end{figure}

\begin{figure}[ht]
\centerline{\psfig{file=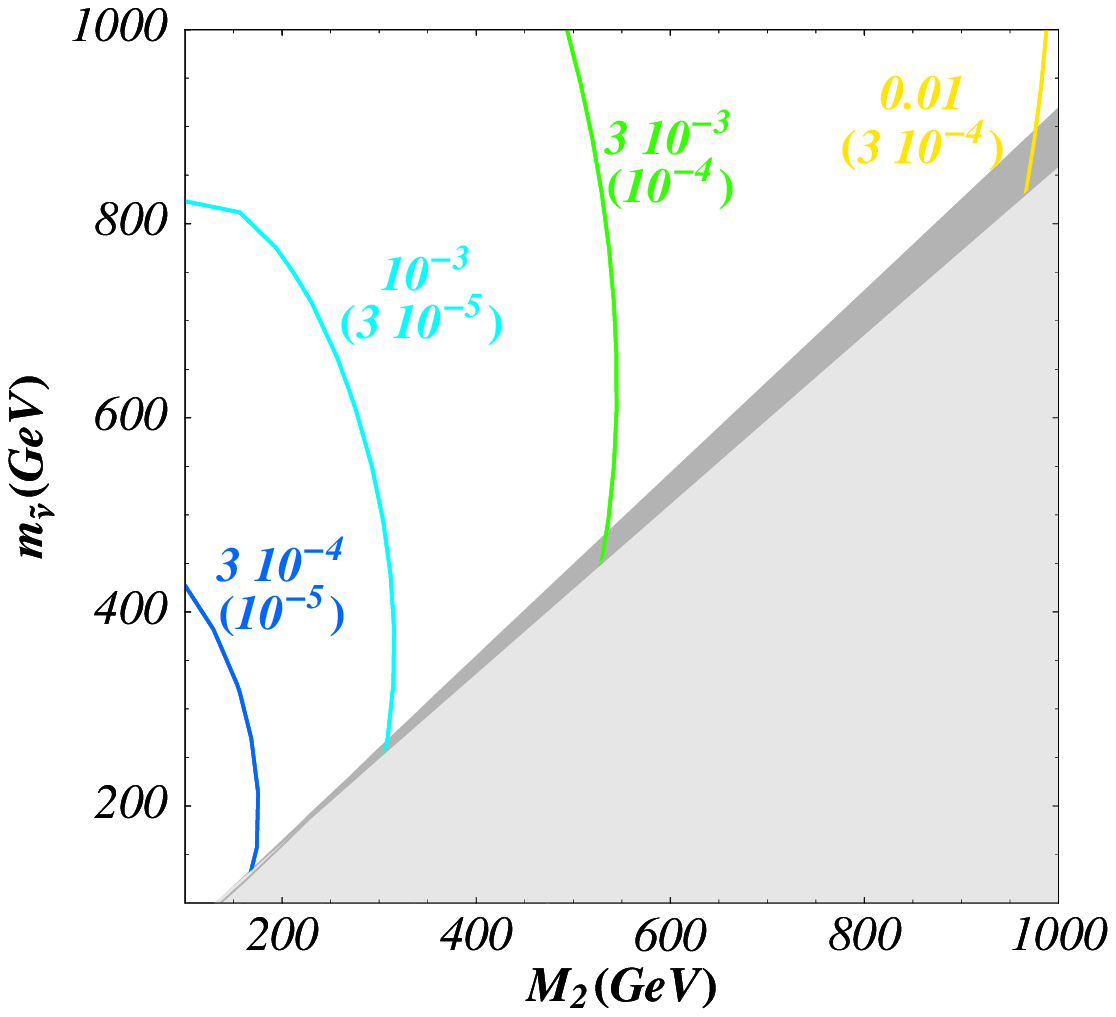,width=0.85\textwidth}}
\caption{Upper bound on $\frac{(m^2_{\tilde \nu})_{e \mu}}{m^2_{\tilde \nu}}$
from the present experimental limit $BR(\mu \rightarrow e \gamma) <
1.2 \times 10^{-11}$ as a function of $m_{\tilde \nu}$ and $M_2,$ for
$\mbox{tan}\beta=10$, $A_0=0$ and $\mbox{sign} (\mu)=+.$
The numbers in brackets correspond to an improvement by a factor $10^3$
in the experimental limit on $BR(\mu \rightarrow e \gamma)$. The regions in
dark grey and light grey are defined as in fig. 1.}
\label{figmueg_1}
\end{figure}

The running of the slepton masses from $M_U$ to the scale where
right-handed neutrinos decouple, due to loop corrections involving the
Yukawa coupling matrix $Y,$ induces flavour non-diagonal terms
in the slepton mass matrices which are proportional to the off-diagonal
entries of
\beq
  C\ =\ Y^\dagger ~\ln \left({M_U/M}\right)~ Y\ .
\label{eq:C}
\eeq
These flavour-dependent
soft terms induce LFV charged leptons radiative decays
$l_i \rightarrow l_j \gamma$ via loops of charginos/neutralinos
and sleptons. The corresponding amplitudes have been fully computed,
in the mass insertion approximation, in Ref. \cite{Hisano99}.
For moderate and large values of $\tan \beta,$ the dominant contribution
arises from loops where charginos and sneutrinos circulate, with an insertion
of the off-diagonal element of the sneutrino mass matrix
${m^2_{\tilde{\nu}}}_{ij}.$ This yields a branching ratio
\beq
  \mbox{BR}\, (l_i \rightarrow l_j \gamma)\ \simeq\ \frac{\alpha^3}{G^2_F}\,
  f (M_2, \mu, m_{\tilde \nu})\, |{m^2_{\tilde L}}_{ji}|^2
  \tan^2 \! \beta\ ,
\label{eq:BR}
\eeq
where $f$ is a function of the $SU(2)_L$ gaugino mass parameter $M_2$,
the supersymmetric Higgs mass $\mu$ and the mean sneutrino mass
$m_{\tilde \nu}$. Thus the dependence of $\mbox{BR} (l_i \rightarrow
l_j \gamma)$ on the seesaw parameters is encoded in the off-diagonal entries
of the matrix $C$; this remains true when the subdominant contributions are
taken into account.
 
In figs. \ref{figtmg_1} and
\ref{figmueg_1}, we show the upper limits on
$(m^2_{\tilde \nu})_{\mu \tau}/m^2_{\tilde \nu}$ and
$(m^2_{\tilde \nu})_{e \mu}/m^2_{\tilde \nu}$
that can be inferred from the present bounds on the branching ratios for
$\tau \rightarrow \mu \gamma$ and $\mu \rightarrow e \gamma$, respectively.
The bounds are displayed in the $(m_{\tilde{\nu}}, M_2)$ plane, the variables
which they are mostly sensitive to, for $\tan \beta = 10,$ $A_0 = 0$ and
$\mbox{sign}\, (\mu) = +.$ However, since
$\mbox{BR} (l_i \rightarrow l_j \gamma)$ scales as $\tan^2 \beta$ for
moderate and large values of $\tan \beta,$ the upper limits associated with
other values of $\tan \beta$ can be obtained upon multiplication by
$(10/\tan\beta)$. We also give in brackets the
limits corresponding to an improvement by three orders of magnitude in the
sensitivity to the branching ratios. Such an improvement is indeed
expected for $\mu \rightarrow e \gamma$ \cite{futdirexp}. Prospects for $\tau
\rightarrow \mu \gamma$ are currently less optimistic, 
but searches for
the LFV decay $\widetilde \chi^0_2 \rightarrow \widetilde \chi^0_1 \mu \tau$
at future colliders could provide limits on 
$(m^2_{\tilde \nu})_{\mu \tau}/m^2_{\tilde \nu}$ 
of that order of
magnitude in the region of the $(m_{\tilde{\nu}}, M_2)$ plane indicated on
figs. \ref{figtmg_1} and \ref{figmueg_1} \cite{Hinchliffe01}. 
We do not
consider the process $\tau \rightarrow e \gamma,$ which is much less
constrained than $\mu  \rightarrow e \gamma$ by experiment, and hence does
not provide additional information on the seesaw parameters.

\begin{figure}[!ht]
\centerline{\psfig{file=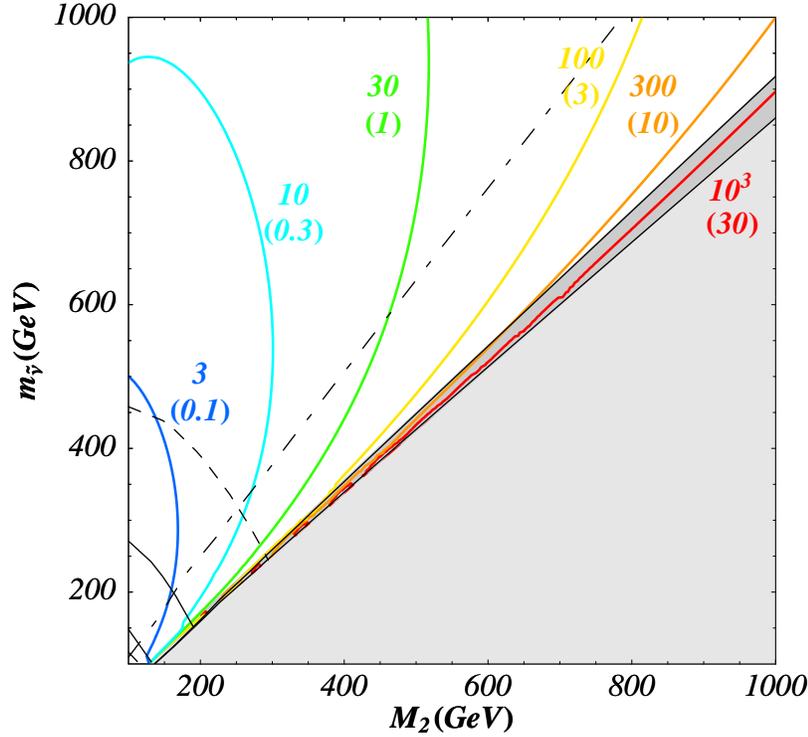,width=0.7\textwidth}}
\caption{Upper bound on $C_{23} =
(Y^\dagger \mbox{ln}(\frac{M_U}{M_R}) Y)_{23}$ from the present
experimental limit $BR(\tau \rightarrow \mu \gamma) < 1.1 \times 10^{-6}$
as a function of the mean sneutrino mass $m_{\tilde \nu}$ and of the
$SU(2)_L$ gaugino soft mass $M_2,$ for $\mbox{tan}\beta=10$, $A_0=0$ and
$\mbox{sign} (\mu)=+.$
The numbers in brackets correspond to an improvement by a factor $10^3$
in the experimental limit on $BR(\tau \rightarrow \mu \gamma)$.
The dark gray region is excluded because the lighter stau would here be the
LSP, while the light grey region corresponds to $m^2_0 < 0$.
The region below the dash-dotted line could be explored by searching for LFV
slepton decays at future colliders \cite{Hinchliffe01}.}
\label{figtmg}
\end{figure}

\begin{figure}[ht]
\centerline{\psfig{file=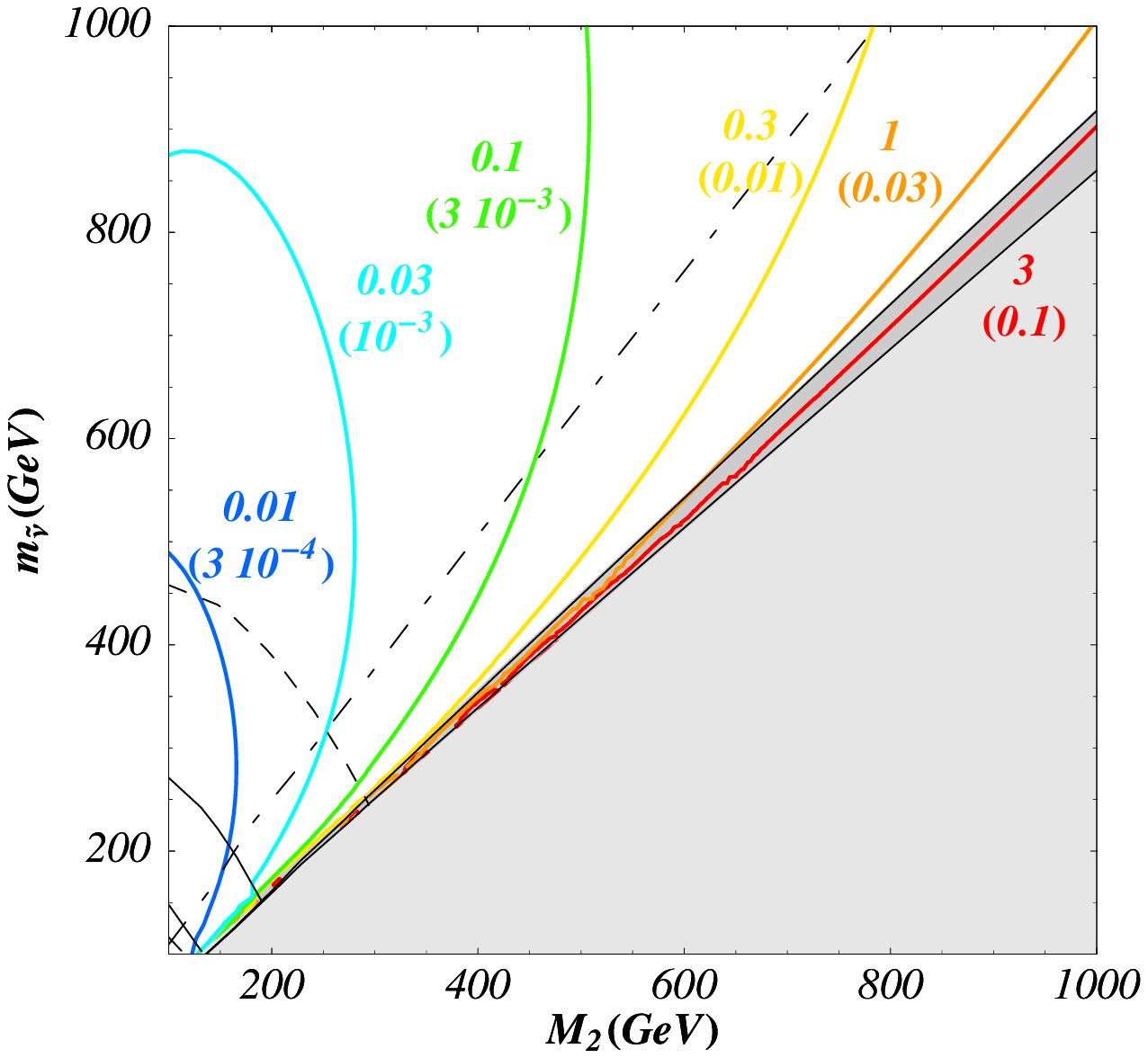,width=0.75\textwidth}}
\caption{Upper bound on
$C_{12}=(Y^\dagger \mbox{ln}(\frac{M_U}{M_R}) Y)_{12}$
from the present experimental limit $BR(\mu \rightarrow e \gamma) <
1.2 \times 10^{-11}$ as a function of $m_{\tilde \nu}$ and $M_2,$ for
$\mbox{tan}\beta=10$, $A_0=0$ and $\mbox{sign} (\mu)=+.$
The numbers in brackets correspond to an improvement by a factor $10^3$
in the experimental limit on $BR(\mu \rightarrow e \gamma)$. The regions in
dark grey and light grey are defined as in fig. 1.}
\label{figmueg}
\end{figure}

In the framework of mSUGRA the coefficients $C_{ij}$ are related 
to the $(m^2_{\tilde \nu})_{\mu \tau}/m^2_{\tilde \nu}$ by 
\beq
 {m^2_{\tilde L}}_{ij}\ \simeq\ -\, \frac{3 m^2_0 + A^2_0}{8 \pi^2}\ C_{ij}\ .
\label{eq:soft_terms}
\eeq
The limits in figs.\ref{figtmg} and \ref{figmueg} corresponds to those in figs.
\ref{figtmg_1} and \ref{figmueg_1}. The latter are taken from \cite{lms1}
where a more complete discussed can be found.

In view of the following discussion, it is useful to express the quantity
$C_{ji}$ in the parametrization (\ref{laL}):
\beq
  C_{ji}\ =\ \frac{z^*_j z_i\, M_z}{v^2}\:
  \ln \left( \frac{M_U}{M_z} \right) +\ \frac{y^*_j y_i\, M_y}{v^2}\:
  \ln \left( \frac{M_U}{M_y} \right) +\ \frac{x^*_j x_i\, M_x}{v^2}\:
  \ln \left( \frac{M_U}{M_x} \right)\ .
\label{eq:C_ij}
\eeq
At this stage, nothing has been assumed about the structure of $L,$
and Eq. (\ref{eq:C_ij}) is completely general. We shall neglect CP
violation in the following, and therefore replace $z^*_j z_i$
by $z_j z_i,$ etc, with an additional $-$ sign if the corresponding
right-handed neutrino has a negative CP eigenvalue.

\subsection{Hierarchical mass spectrum}

In the case of a hierarchical light neutrino mass spectrum,
$m_@ \approx z^2_+,$ and it is convenient to rewrite Eq. (\ref{eq:C_ij}) as
\beq
  C_{ji}\ =\ \frac{z^*_j z_i}{z^2_+}\ \frac{M_z}{M_@}\:
  \ln \left( \frac{M_U}{M_z} \right) +\ \frac{y^*_j y_i}{z^2_+}\
  \frac{M_y}{M_@}\: \ln \left( \frac{M_U}{M_y} \right) +\
  \frac{x^*_j x_i}{z^2_+}\ \frac{M_x}{M_@}\:
  \ln \left( \frac{M_U}{M_x} \right)\ ,
\label{eq:C_hierarchy}
\eeq
where $M_@ \equiv v^2 / m_@ \sim 5 \times 10^{14}$ GeV.
The dominance hypothesis, together with the assumed perturbativity of Yukawa
couplings, implies $M_z \lesssim M_@.$
In the case of double dominance, one also has $M_y \lesssim M_{\odot}
\equiv v^2 / m_{\odot}$ (with typical values $M_{\odot} \sim 5 \times 10^{15}$
GeV for LMA, and $M_{\odot} \sim 10^{17}$ GeV for LOW), and
$M_x \lesssim M_1 \equiv v^2 / m_1$ ($M_1 \gg M_{\odot}$).
Since $m_1$ is not known, the constraint on $M_x$ is not very informative;
however if one gives up the dominance in the solar neutrino sector, one
has both $M_x \lesssim M_{\odot}$ and $M_y \lesssim M_{\odot}.$
In the pseudo-Dirac case, Eq. (\ref{eq:C_ij}) further simplifies to
\beq
  C_{ji}\ \simeq\ \frac{z_j z_i}{z^2_+}\ \frac{M_z}{M_@}\:
  \ln \left( \frac{M_U}{M_z} \right) +\ \frac{y_j y_i - x_j x_i}{z^2_+}\
  \frac{\bar M}{M_@}\: \ln \left( \frac{M_U}{\bar M} \right)\ .
\label{eq:C_pseudo}
\eeq
Note that the $\ln (M_U / \bar M)$ term in Eq. (\ref{eq:C_pseudo}) is
proportional to $y_j y_i - x_j x_i,$ {\it not} to $y_j y_i + x_j x_i.$ This
is important since cancellations may occur in one of the two combinations,
depending on how the pseudo-Dirac structure is realized.
More precisely, one may have $|y_1 y_2 - x_1 x_2| \ll |y_1 y_2 + x_1 x_2|$
if cancellations occur both in $m_{11}$ and $m_{22}$ (implying a suppression
of the contribution of the pseudo-Dirac right-handed neutrino pair to
$C_{12}$), and $|y_2 y_3 + x_2 x_3| \ll |y_2 y_3 - x_2 x_3|$ if cancellations
occur in $m_{22}.$ Furthermore one has $\bar M \lesssim v^2 / \bar m \simeq
\sqrt{2 |1 - \tan^2 \theta_{12}|} M_{\odot}.$
Note however that the actual upper bound on $\bar M$ can be smaller, even in
the presence of Yukawa couplings of
order one. In fact $\bar M \sim \sqrt{2 |1 - \tan^2 \theta_{12}|}\ M_{\odot}$
requires, in addition to Yukawa couplings of order one, both cancellations in
$m_{11}$ and $m_{22},$ and $|y_1| \sim |y_-|$).

\subsubsection{$\tau \rightarrow \mu \gamma$}

In the double dominance case, the coefficient associated with the decay
$\tau \rightarrow \mu \gamma$ is
\beq
  C_{23}\ \simeq\ \frac{1}{2} \sin 2 \theta_{23}\ \frac{M_z}{M_@}\:
  \ln \left( \frac{M_U}{M_z} \right)\ +\ \frac{y_2 y_3}{z^2_+}\
  \frac{M_y}{M_@}\: \ln \left( \frac{M_U}{M_y} \right)\
  +\ \frac{x_2 x_3}{z^2_+}\ \frac{M_x}{M_@}\:
  \ln \left( \frac{M_U}{M_x} \right)\ .
\label{eq:C_23_double}
\eeq
Although, from Eq. (\ref{ddrelations}), $|y_2 y_3| \lesssim m_{\odot}$ and
$|x_2 x_3| \ll m_{\odot},$ the second and third terms in Eq.
(\ref{eq:C_23_double}) can be as large as
$\ln (M_U / M_{\odot})$ and $\ln (M_U / M_1),$ respectively, if $M_y$
and $M_x$ are close to their respective upper bounds.
In the pseudo-Dirac case:
\beq
  C_{23}\ \simeq\ \frac{1}{2} \sin 2 \theta_{23}\ \frac{M_z}{M_@}\:
  \ln \left( \frac{M_U}{M_z} \right)\ +\ \frac{y_2 y_3 - x_2 x_3}{z^2_+}\
  \frac{\bar M}{M_@}\: \ln \left( \frac{M_U}{\bar M} \right)\ .
\label{eq:C_23_pseudo}
\eeq
In general one expects the contribution of the pseudo-Dirac pair of
right-handed neutrinos to be small, since from Eq. (\ref{eq:PD_hierarchy})
$|y_2 y_3 + x_2 x_3| \ll m_{\odot} / \sqrt{2 |1 - \tan^2 \theta_{12}|}.$
Still it can be of order one when cancellations occur in $m_{22}$ due to
both $y_3 \simeq \pm i x_3$ and $y_2 \simeq \pm i x_2,$ since in this
case $|y_2 y_3 - x_2 x_3| \gg |y_2 y_3 + x_2 x_3|.$

If $M_z$ is the heaviest right-handed neutrino mass, Eqs.
(\ref{eq:C_23_double}) and (\ref{eq:C_23_pseudo}) reduce to $C_{23}
\simeq \frac{1}{2} \sin 2 \theta_{23}\, \frac{M_z }{M_@} $ $\ln (M_U / M_z);$
in this case the observation of the process $\tau \rightarrow \mu \gamma$
would amount to a measurement of $M_z$ as a function of $M_2$ and
$m_{\tilde \nu}$ -- assuming no other contribution to the LFV slepton masses,
\eg\ from the supersymmetry breaking mechanism itself. In general, however,
$M_z$ can be smaller than $M_x$ or $M_y$ (resp. $\bar M$), and
the second or third term may give the dominant contribution to
$C_{23}.$ Hence a measurement of $\mbox{BR} (\tau \rightarrow \mu \gamma)$
can only provide an upper bound on $M_z.$ A small branching ratio,
$|C_{23}| \ll \sin 2 \theta_{23} \ln (M_U / M_@),$ would point towards two
alternative scenarios \cite{lms1}:
(i) a scenario characterized by small Yukawa couplings
in the neutrino sector and a ``low'' lepton number breaking scale,
with both $M_z \ll M_@$ and $M_x, M_y \ll M_{\odot}$ (resp.
$\bar M \ll \sqrt{2 |1 - \tan^2 \theta_{12}|}\, M_{\odot}$);
(ii) a scenario characterized by the dominance of a lighter right-handed
neutrino in the atmospheric sector and a ``high'' lepton number breaking
scale, with $M_z \ll M_@$ and $M_y \sim M_{\odot}$ or $M_x \gtrsim M_{\odot}$
(resp. $\bar M$ close to $\sqrt{2 |1 - \tan^2 \theta_{12}|}\, M_{\odot}$),
the contribution of $M_y$ (resp. $\bar M$) to $C_{23}$ being suppressed by
$|x_2 x_3|, |y_2 y_3| \ll m_{\odot}$ (resp. $|y_2 y_3 - x_2 x_3| \ll
m_{\odot} / \sqrt{2 |1 - \tan^2 \theta_{12}|}$). In the latter, the $B-L$
symmetry is broken close to -- or even above -- the scale where the MSSM gauge
couplings unify (\ie\ $2 \times 10^{16}$ GeV), which favours grand unified
theories based on $SO(10)$ and larger gauge groups. On the contrary the former
appears to disfavour such theories.

\subsubsection{$\mu \rightarrow e \gamma$}

Let us first discuss the LMA and LOW solutions, which are strongly favoured
over SMA by the solar neutrino data. In the double dominance case, the
coefficient associated with the decay $\mu \rightarrow e \gamma$ is
\beq
  C_{12}\ \simeq\ \frac{z_1}{z_+}\, \sin \theta_{23}\, \frac{M_z}{M_@}\:
  \ln \left( \frac{M_U}{M_z} \right) +\ \frac{y_1 y_2}{z^2_+}\
  \frac{M_y}{M_@}\: \ln \left( \frac{M_U}{M_y} \right)\
  +\ \frac{x_1 x_2}{z^2_+}\ \frac{M_x}{M_@}\:
  \ln \left( \frac{M_U}{M_x} \right)\ .
\label{eq:C_12_hierarchy}
\eeq
In the pseudo-Dirac case,
\beq
  C_{12}\ \simeq\ \frac{z_1}{z_+}\, \sin \theta_{23}\, \frac{M_z}{M_@}\:
  \ln \left( \frac{M_U}{M_z} \right) +\ \frac{y_1 y_2 - x_1 x_2}{z^2_+}\
  \frac{\bar M}{M_@}\: \ln \left( \frac{M_U}{\bar M} \right)\ .
\label{eq:C_12_pseudo}
\eeq
One can show that the contribution of the pseudo-Dirac pair of right-handed
neutrinos is always suppressed relative to
$\ln\, (M_U / M_{\odot} \sqrt{2 |1 - \tan^2 \theta_{12}|}\, ),$ due to either
$|y_1 y_2 - x_1 x_2| \ll m_{\odot} / \sqrt{2|1 - \tan^2 \theta_{12}|}$ or
$\bar M \ll \sqrt{2 |1 - \tan^2 \theta_{12}|}\, M_{\odot},$ while in the
double dominance case $M_y$ or $M_x$ could in principle give a contribution
of order one to $C_{12}.$

Expectations for $\mbox{BR} (\mu \rightarrow e \gamma)$ are very
model-dependent. It is useful for the discussion to distinguish between two
regimes differing by the value of $|U_{e3}|.$ In the ``large $|U_{e3}|$''
regime, characterized by $2 |U_{e3}| \gg \sin 2 \theta_{12}\, m_{\odot}
/ m_@$ in the double dominance case (resp. $|U_{e3}| \gg m_{\odot} /
\sqrt{2 |1 - \tan^2 \theta_{12}|}\, m_@$ in the pseudo-Dirac case, where this
regime exists only for the LOW solution), the first term in Eqs.
(\ref{eq:C_12_hierarchy}) and (\ref{eq:C_12_pseudo}) yields a $M_z$-dependent
lower bound on $|C_{12}|,$
\beq
  |C_{12}|\ \gtrsim\ \sin \theta_{23} \tan \theta_{13}\, \frac{M_z}{M_@}\:
  \ln \left( \frac{M_U}{M_z} \right)\ .
\label{eq:C_12_double}
\eeq
Eq. (\ref{eq:C_12_double}) can be used to convert an experimental limit on
$\mbox{BR} (\mu \rightarrow e \gamma)$ into an upper bound on $M_z$ as a
function of $M_2$ and $m_{\tilde \nu}$ -- assuming $|U_{e3}|$ is known.
Actually the present limit on $\mbox{BR} (\mu \rightarrow e \gamma)$ already
provides strong constraints on this regime; in particular a large $M_z,$
$M_z \sim M_@,$ is already excluded over a large portion of the
($m_{\tilde \nu},$ $M_2$) plane,
especially for LMA. The contribution of the other right-handed neutrinos
also yields non-trivial constraints. In the double dominance case,
$|C_{12}| \ll 1$ requires $|y_1 y_2| \ll \sin 2 \theta_{12} m_{\odot} / 2$
and/or $M_y \ll M_{\odot}$ (there are no analogous constraints for $|x_1 x_2|$
and $M_x$ since $m_1$ is unknown). In the pseudo-Dirac case,
$|C_{12}| \ll 1$ requires $|y_1 y_2 - x_1 x_2| \ll
m_{\odot} / \sqrt{2 |1 - \tan^2 \theta_{12}|}$
and/or $\bar M \ll \sqrt{2 |1 - \tan^2 \theta_{12}|}\, M_{\odot},$
which is easily obtained due to the pseudo-Dirac structure.
With the expected improvement by three orders of magnitude
in the experimental limit on $\mbox{BR} (\mu \rightarrow e \gamma),$ those
constraints should become rather stringent -- unless $\mu \rightarrow e \gamma$
is observed.

Note finally that if atmospheric neutrino oscillations are dominated by the
heaviest right-handed neutrino mass (\ie\ $M_x, M_y \leq M_z,$ resp.
$\bar M \leq M_z$), the contribution of $M_z$ dominates both in $\mbox{BR}
(\tau \rightarrow \mu \gamma)$ and in $\mbox{BR} (\mu \rightarrow e \gamma)$.
This leads to the prediction
\beq
  \frac{|C_{12}|}{|C_{23}|}\ \simeq\ \frac{\tan \theta_{13}}
  {\cos \theta_{23}}\ .
\label{eq:C_12_23_double}
\eeq
While Eq. (\ref{eq:C_12_23_double}) remains valid over a large region of the
seesaw parameter space with $M_x > M_z$ or $M_y > M_z$ (resp. $\bar M > M_z$),
a deviation from this relation is possible only if $M_z$ is not the largest
right-handed neutrino mass. In this case one may have $|C_{12}| \gg |U_{e3}|
\ln\, (M_U / M_@) \approx 8\, |U_{e3}|$ (which is not excluded for the LOW
solution yet), hence an observable rate for $\mu \rightarrow e \gamma.$

The ``small $|U_{e3}|$'' regime, characterized by
$2 |U_{e3}| \sim \sin 2 \theta_{12}\, m_{\odot} / m_@$ in the
double dominance case (resp. $|U_{e3}| \sim m_{\odot} /
\sqrt{2 |1 - \tan^2 \theta_{12}|}\, m_@$ in the pseudo-Dirac case),
differs from the large $|U_{e3}|$ regime by the absence of a lower
bound on $|C_{12}|$: here $z_1 \rightarrow 0$ is perfectly
compatible with the existing data. Thus $\mbox{BR} (\mu \rightarrow e \gamma)$
can be extremelly small even though $M_z \sim M_@$ (\ie\ even though
$\mbox{BR} (\tau \rightarrow \mu \gamma)$ is large), unlike in the previous
regime.
However $|C_{12}| \gg |U_{e3}| \ln (M_U/M_@)$ is also a possibility when
$M_x$ or $M_y \gg M_@ \gtrsim M_z$ (resp. $\bar M \gg M_@ \gtrsim M_z$),
and $\mu \rightarrow e \gamma$ should be observed at PSI in this case.

The discussion of the (strongly disfavoured) SMA solution follows the same
lines. Again one can distinguish between two extreme regimes. In the
``large $|U_{e3}|$'' regime, where $|U_{e3}| \sim \tan \theta_{12},$
the lower bound (\ref{eq:C_12_double}) holds like for the large angle
solutions, as well as Eq. (\ref{eq:C_12_23_double}) in a large portion of
the seesaw parameter space including the case where $M_z$ is the largest
right-handed neutrino mass. Note however that
$\mbox{BR} (\mu \rightarrow e \gamma)$ cannot be as large as in the LMA
or LOW case, due to the upper bound $|C_{12}| \lesssim \tan \theta_{12}
\ln (M_U / M_@)$ (to be compared with $|C_{12}| \lesssim
\ln (M_U / M_{\odot})$ for LMA/LOW). In the ``small $|U_{e3}|$'' regime,
where $|U_{e3}| \sim \tan \theta_{12}\, m_{\odot} / m_@,$
$\mbox{BR} (\mu \rightarrow e \gamma)$ can be extremelly small without
conflicting with existing data. Again the observation of
$\mu \rightarrow e \gamma$ with $|C_{12}| \gg |U_{e3}| \ln (M_U/M_@)$
is possible if one of the two lightest right-handed neutrinos dominates
in the atmospheric sector, \ie\ if $M_x$ or $M_y \gg M_@ \gtrsim M_z.$


\subsection{Inverted Hierarchy Mass Spectrum} \label{sec:iHlfv}

When the neutrino mass matrix is endowed with an approximate pseudo-Dirac
pattern as required by an inverted hierarchy in its eigenvalues, it is
convenient to rewrite the coefficients $C_{ij}$ given by (\ref{eq:C_hierarchy})
in terms of the variables defined in Section \ref{sec:iHsolar} as follows: 
\bea
  C_{ji}\ & = &\ \frac{\left(u^*_j u_i + v^*_j v_i \right)}{m_@} 
  \ \frac{\Md}{M_@}\: \ln \left( \frac{M_U}{\Md} \right)  +\
  \frac{x^*_j x_i}{u_+ v_1}\ \frac{M_x}{M_@}\:
  \ln \left( \frac{M_U}{M_x} \right) \cr
{} & = &\ \left(\frac{u^*_j u_i}{u^2_+} e^{2\lambda} + 
\frac{v^*_j v_i}{v^2_1} e^{-2\lambda}\right) \ \frac{\Md}{M_@}\:
\ln \left( \frac{M_U}{\Md} \right)  +\
\frac{x^*_j x_i}{m_@}\ \frac{M_x}{M_@}\: 
\ln \left( \frac{M_U}{M_x} \right)\ ,
\label{eq:C_iH}
\eea
Since, as discussed, an approximate pseudo-Dirac texture in the
neutrino mass matrix is more naturally realized by a corresponding
pseudo-Dirac pair of right-handed neutrinos, we have associated the
masses $\pm \Md$ to this pair.  The dependence on the rapidity
$\lambda$ shows the dependence on the boosts (\ref{boost}) and
displays an example of the dependence of the $C_{ij}$ on the orthogonal
tranformations discussed in Section \ref{sec:seesaw} .

The Yukawa coupling associated to the largest entry in the
matrix $L$, which in this case must be either $u_+$ or $v_1$,
is perturbative if $ \Md \lesssim e^{-2|\lambda |} M_@ \ .$
The best limit on $M_x$ is obtained when the $x_i$ contributions 
saturate the quantities given by (\ref{psDit13}) and (\ref{psDiapprox}):
$M_x \lesssim (m_{\odot}^2 /2U^2_{e3} m^2_@ )M_@ \ \lesssim M_@ / |U_{e3}|
\, ,$ where the last bound relies on (\ref{psDiUe3}), so that $M_x$
lies well above $ M_@.$

\subsubsection{$\tau \rightarrow \mu \gamma$}

The coefficient associated to the $\tau \rightarrow \mu \gamma$ decay
is
\beq
C_{23} \approx - \frac{1}{2} \sin 2\theta_{23} e^{2\lambda}\
\frac{\Md}{M_@}\: \ln \left( \frac{M_U}{\Md} \right) +
\frac{ x^2_+ +x^2_- }{2m_@} \sin 2(\theta_{23} + \gamma)
\frac{M_x}{M_@}\: \ln \left( \frac{M_U}{M_x} \right)\ ,
\label{eq:C_23}
\eeq
where $\tan \gamma = x_+ / x_-\, .$ Since $\sin 2\theta_{23}$ is large,
the first term gives a contribution ${\cal O}(1)$ when $ \Md $
approaches its bound $ e^{-2|\lambda |} M_@ \ $, which could be much
below the gauge coupling unification scale if $\lambda > 1\, .$ On the
contrary, this contribution is suppressed by a factor $ e^{-2|\lambda |}$ for
$\lambda < 0\, .$ As for the second term, it is strongly suppressed by
the bounds on the $x_i$ factors, such that $(x^2_+ +x^2_- )/{2m_@}$ is
expected to be ${\cal O}({m_{\odot}^2}/{2m^2_@})\, $. This suppression
can be compensated if the mass $M_x$ gets close to the bounds imposed
by the Yukawa coupling perturbativity condition. For LMA the bound on
$M_x$ lies near the gauge coupling unification scale, but for the more
appropriate LOW solution it is around the Planck scale, which is
clearly nonsense. Therefore, some suppression is expected on quite
general grounds for this term.

\subsubsection{$\mu \rightarrow e \gamma$}

The coefficient $C_{12}$ corresponding to the $\mu \rightarrow 
e \gamma$ decay is
\bea
C_{12} & \approx & \left[ \left( \frac{u_1}{u_+} e^{2\lambda} + 
\frac{v_+}{v_1} e^{-2\lambda}\right) \cos \theta_{23} +
\frac{v_-}{v_1} e^{-2\lambda} \sin \theta_{23} \right]
\frac{\Md}{M_@}\: \ln \left( \frac{M_U}{\Md} \right) \cr
& + & \frac{x_1 \sqrt{x^2_+ +x^2_-}}{m_@} \sin (\theta_{23} +\gamma )
\frac{M_x}{M_@}\: \ln \left( \frac{M_U}{M_x} \right)\ ,
\eea

For simplicity, let us separately consider the two typical situations
previously identified in Section \ref{sec:iHsolar} :

\noindent 1) {\it Double dominance by a pseudo-Dirac pair;} replacing
the ratios $(u_1 /u_+ ),\, (v_+ /v_1 ),\, (v_- /v_1 ) ,\,$ by the
expressions obtained from (\ref{psDit13}) and (\ref{psDiapprox}), yields
\bea
C_{12} & \approx & 
\left[ \left( \frac{m_{\odot}^2}{4m^2_@}\cosh{2\lambda} + 
(\tan \theta_{12} - 1)\sinh {2\lambda}\right) \cos \theta_{23} -
 e^{-2\lambda} \tan{\theta_{13}}\sin \theta_{23} \right]
\frac{\Md}{M_@}\: \ln \left( \frac{M_U}{\Md} \right) \cr
& \sim & {\cal O} \left( \frac{m_{\odot}^2}{4m^2_@}e^{2\lambda}
\frac{\Md}{M_@}\: \right)  \ln \left( \frac{M_U}{\Md} \right) 
 \lesssim  (8 + 2|\lambda |)\frac{m_{\odot}^2}{4m^2_@}
\eea
where the last evaluations corresponds to the generic case - \ie , to a
natural pattern in the sense adopted in this paper - and $\lambda >0\,
.$ This limit is very small for the LOW case, the one most relevant
here.  Notice that in this case, the approximate relation holds: 
$
C_{12} / C_{23} \sim {\cal O}\left( {m_{\odot}^2}/{2m^2_@}\right)
\label{12/23}
$
The upper limit on $C_{12}$ can be saturated only if $\Md$ is close to
its bound which implies a relatively large $\tau \rightarrow \mu
\gamma$ decay. Of course, this contribution can be overcome by the
other one if $M_x$ is very large.  

\noindent 2) {\it Single dominance of} $m_{\odot}\, ;$ then the
components $x_i$ can be evaluated from (\ref{psDit13}) and
(\ref{psDiapprox}) in terms of $m_{\odot},\, (\tan^2 \theta_{12} - 1),\,
U_{e3}.$ However, the generic relation: $x_1 x_- \sim x_1 x_+
\lesssim (x_1^2 +x_+^2)/2 \approx {m_{\odot}^2}/{4m_@} $ leads to the
estimate 
\beq
C_{12} \sim {\cal O}\left(\frac{m_{\odot}^2}{4m^2_@}\right)
\frac{M_x}{M_@}\: \ln \left( \frac{M_U}{M_x} \right) \, . 
\eeq
Therefore, in this case, $C_{12}$ could be of ${\cal O}(1)$ only if
$M_x$ lies close to its bound $M_@ / |U_{e3}|\, .$ However, for the more
suitable LOW solution, (\ref{psDiUe3}) implies that $M_x$ is too close
(or above) $M_U$ for the result to be sensible.  By assuming that the
contributions from the $x_i$ dominate in both $C_{12}$ and $C_{23}$
their ratio is given by:
$ {C_{12}}/{C_{23}} \approx {x_1}/
{\sqrt{x^2_+ +x^2_-}\cos (\theta_{23} +\gamma )} ,  $
which is less suppressed than in the previous case and 
just tells us that the right-handed
neutrino with mass $M_x$ is more coupled to $e$ than to $\mu$ or
$\tau$.  This is not possible in the previous case of double dominance
by a pseudo-Dirac pair because this ratio is suppressed by the
conditions to implement a pseudo-Dirac pair. Of course the important
prediction from the experimental viewpoint is for each coefficient.


\section{On the Physical Interpretation of $R$}\label{sec:R}

The ambiguities in the extraction of the see-saw parameters from the
measurements of $\cal{M}_\nu$ can be expressed  \cite{Casas01} in
terms of the complex orthogonal matrix $R$ defined by:
\beq
R  \left( \matrix{ \sqrt{m_1} &  & \cr 
                      & \sqrt{m_2} &  \cr  
                  &  & \sqrt{m_3} }\right) U^{\dagger}  = M^{-1/2}  Y v = L  
\label{Rdef}
\eeq
The matrices are defined as in the previous sections. The nice property
of this choice of $R$ is its immediate interpretation as the
dominance matrix in the sense that: \\
\noindent {\sl (i)} $R$ is an orthogonal
tranformation from the basis of the left-handed leptons mass
eigenstates to the one of the right-handed neutrino mass eigenstates;\\
\noindent ({\sl ii}) if and only if an eigenvalue $m_i$ of ${\cal M}_\nu$ is
dominated - in the sense already given before - by one right-handed
neutrino eigenstate $N_j$, then $|R_{ji}| \approx 1\, ;$\\
\noindent {\sl (iii)} if a light pseudo-Dirac pair is dominated by a heavy pseudo-Dirac pair,
then the corresponding  $2 \times 2$ sector in $R$ is a boost.

It is instructive to illustrate this property by a simple example,
which in a sense also motivates the double dominance assumption in
Section \ref{sec:solar}. Consider a hierarchical neutrino mass
spectrum, $m_3 \approx m_@ \gg m_2 \approx m_{\odot} \gg m_1\, .$ For the
sake of the example, we fix the other 9 parameters (all phases are
neglected) as follows: ({\sl a}) The $Y$ eigenvalues are very
hierachical with $y_3 \gg y_2  \gg y_1\, ;$ ({\sl b}) bimaximal mixing
angles in $U\, ;$ ({\sl c}) the $V_L$ matrix defined, as usual, by the
diagonalization $ Y = V_R^{\dagger} \mbox{diag}(y_1,\, y_2,\, y_3) V_L, $ is
very small or very close to the deviations of $U$ from the bimaximal
case. This last assumption is interesting because it selects the case
where the charged lepton flavour violations are very small, while the
mixings are maximal in ${\cal M}_\nu \, .$ Then we get the approximate
expressions for the $M$ eigenvalues:
\beq
M_1 = \frac{2y_1^2}{m_{\odot}} \quad M_2 = \frac{2y_2^2}{m_{@}} 
\quad M_3 = \frac{y_3^2}{4m_{1}} \, , \label{example}
\eeq
where $M_3$ is much larger than $M_2 \ll M_{@}$, while $M_1 \ll
M_{\odot}$ depends on the hierarchy in the $Y$ eigenvalues. Notice that
our assumption of hierarchical spectrum for the left-handed neutrinos
and small charged lepton flavour violations leaded to a pattern of $M$
eigenvalues of the kind discussed in \cite{lms1}.

As for the dominance matrix $R$, which in this case is a rotation matrix, 
we get
\beq
R \approx \left( \matrix{ \sqrt{\frac{m_1}{m_{\odot}}} & 1 & 
\sqrt{\frac{m_{\odot}}{m_@}}\cr 
-3 \sqrt{\frac{m_1}{2m_@}} & -\sqrt{\frac{m_{\odot}}{m_@}} & 1 \cr  
1 & -\sqrt{\frac{m_1}{m_{\odot}}} & \sqrt{\frac{2m_1}{m_@}} }\right)~~~~~.
\eeq
The entries (close to) $1$ in $R$ clearly indicate that $m_@$ is dominated
by $N_2 ,\ m_{\odot}$ by $N_1$ and, of course, $m_1$ by $N_3 \, .$
In comparing with the discussion in Ref. \cite{lms1}, one should
take into account the different ordering of the right-handed neutrinos
(by increasing masses) therein. It is worth stressing that our assumptions
of hierarchical light neutrinos and $Y$ eigenvalues {\it yielded} the double
dominance pattern in (\ref{example}). Analogous exercises show how the
dominance of a pseudo-Dirac pair displays in $R$. 

As a second, more familiar example, we consider the class of models
already mentioned, in which the fermion mass matrices are explained by
an abelian flavour symmetry with only positive charges for all leptons.
It is well known that in this case  ${\cal M}_\nu$ is independent of
the right-handed neutrino charges and the magnitude of its matrix
elements $({\cal M}_\nu)_{ij}$ are of ${\cal O}(\ep ^{\l _i + \l
_j})$.  It is easily seen that, in this framework, the $R$ matrix
elements come out all of ${\cal O}(1)\, ,$ characterizing the fact that
all heavy neutrino states are comparably contributing to all light
neutrino mass eigenvalues.

\section{Outlook and Conclusions}

To have a glimpse at the large solar mixing angle from a bottom-up perspective,
we adopt the requirements of robustness and economy and we study the patterns 
for $Y$ and $M$ which, in this sense, naturally account for neutrino phenomenology.
A quantitative appraisal of the structure of subleading terms and of the effect of CP
violating phases, is done by means of reliable approximate expressions. 
Three scenarios, which admit a suggestive physical interpretation in terms of the role
played by right-handed neutrinos, are favoured:
single dominance in the atmospheric sector accompanied by single or pseudo-Dirac dominance in the 
solar sector, in the case of a hierarchical neutrino mass spectrum;  
pseudo-Dirac dominance in the atmospheric sector in the case of an inverted hierarchical spectrum.
 
The phenomenology of the three scenarios is quite rich.
In the hierarchical case, experimentally relevant lower bounds for $U_{e3}$ are derived. 
In the inverted hierarchical case, 
on the contrary, upper bounds of the order of permille are obtained.
All three scenarios predict a rate for $\beta \beta 0 \nu$ too small to be observed
(for a recent exhaustive analysis see Ref. \cite{fsv}).
Connections with the rate of lepton flavour violating decays
$\tau \rightarrow \mu \gamma$ and  $\mu \rightarrow e \gamma$ have also been discussed.
These rates are linked to the nine hidden parameters of the see-saw.
The strong dependence on the three right-handed eigenvalues is manifest,
but also the parameters belonging to the $R$ matrix play an important role. 
For instance, in the pseudo-Dirac dominance scenario, there is a
very strong dependence on the 'rapidity' defined in Eq. (\ref{boost}), which represents
the magnitude of one of the boosts contained in $R$. 

It is worth emphasizing that, while being 'minimalist' from the bottom-up perspective, 
these favourite scenarios are inconsistent with a simple flavour symmetry like a $U(1)$ with only
positive charges for leptons. Indeed, they suggest the presence of much richer flavour symmetries, 
like $U(1)$'s with holomorphic zeros from supersymmetry \cite{afm1, bizarre} or non-abelian flavour 
symmetries \cite{u2}. 
Finally, we also stress that the matrix $R$ \cite{Casas01} admits
a suggestive physical interpretation. Namely, it encodes the informations on the
dominance mechanism at work. 

\vskip 1cm
\noindent
{\bf Note added in proof:}
As stressed in the Introduction, the different dominance scenarios 
analysed in this paper from a bottom-up perspective
have been discussed several times in the literature, 
mostly from a top-down viewpoint.   
The dominance of one right-handed neutrino first appeared in Ref. \cite{Smir93} 
and both the single and double dominance mechanisms consistent with large mixing angles
were considered in Refs. \cite{King99, afm1}. The dominance of a pseudo-Dirac pair in the solar sector
was previously displayed in Refs. \cite{barb98, King99}. Pseudo-Dirac dominance in the atmospheric 
sector has been recently discussed in Refs. \cite{iHbreak}.


\vskip 1cm
\noindent
{\bf Acknowledgements:}
this work has been supported in part by the RTN European Program
HPRN-CT-2000-00148.




\begin{thebibliography}{99}

\bibitem{fit_including_SNO}
G.~L.~Fogli, E.~Lisi, D.~Montanino and A.~Palazzo, Phys. Rev. {\bf D64} (2001) 093007,
hep-ph/0106247;
J.~N.~Bahcall, M.~C.~Gonzalez-Garcia and C.~Pena-Garay, JHEP {\bf 0108} (2001) 014,
hep-ph/0106258. 

\bibitem{SNO} SNO collaboration, Phys. Rev. Lett. {\bf 87} (2001) 71301, nucl-ex/0106015.

\bibitem{solar}  Super-Kamiokande Collaboration, Phys. Rev. Lett. {\bf 86} (2001) 5656, 
hep-ex/0103033.

\bibitem{atmlarge}  Super-Kamiokande Collaboration, Y.~Fukuda {\it et al.}, 
Phys.\ Rev.\ Lett.\ {\bf 81} (1998) 1562, Phys.\ Rev.\ Lett.\ {\bf 85} (2000) 3999.

\bibitem{CHOOZ} M.~Apollonio {\it et al.}, Phys.\ Lett.\ {\bf B420} (1998) 397.

\bibitem{seesaw} M.~Gell-Mann, P.~Ramond and R.~Slansky, in {\it Supergravity}, ed. P. van Nieuwenhuizen 
and D.Z. Freedman, North-Holland, Amsterdam, 1979; T.~Yanagida,  in {\it Proceedings of the Workshop on 
unified theory and baryon number in the universe}, ed. O.~Sawada and A.~Sugamoto, Tsukuba, Japan, 1979;
R.~Mohapatra and G.~Senjanovic, Phys.\ Rev.\ Lett.\ {\bf 44} (1980) 912.

\bibitem{tau_mu_gamma}
CLEO Collaboration, S.~Ahmed {\it et al.},
Phys.\ Rev.\ {\bf D61} (2000) 071101.

\bibitem{mu_e_gamma}
MEGA Collaboration, M.~L.~Brooks {\it et al.},
Phys.\ Rev.\ Lett.\ {\bf 83} (1999) 1521.

\bibitem{isaIJMPA}  
I. Masina, Int. J. Mod. Phys. {\bf A16} (2001) 5101, hep-ph/0107220.

\bibitem{Irges98} N.~Irges, S.~Lavignac and P.~Ramond, Phys.\ Rev.\ {\bf D58} (1998) 035003;
J.~K.~Elwood, N.~Irges and P.~Ramond, Phys.\ Rev.\ Lett.\  {\bf 81} (1998) 5064;
F. Vissani, JHEP {\bf 9811} (1998) 025, hep-ph/9810435;
K. S. Babu, J. C. Pati and F. Wilczek, Nucl. Phys. {\bf B566} (2000) 33, hep-ph/9812538;
F. Vissani, Phys. Lett. {\bf B508} (2001) 79, hep-ph/0102236.

\bibitem{Smir95} A. Yu. Smirnov, Nucl.Phys. {\bf B466} (1996) 25, hep-ph/9511239.
 
\bibitem{King99} S.~F.~King, Nucl.\ Phys.\ {\bf B562} (1999) 57, hep-ph/9904210; 
Nucl. Phys. {\bf B576} (2000) 85, hep-ph/9912492. 

\bibitem{afm1} G.~Altarelli, F.~Feruglio and I.~Masina, Phys.\ Lett.\ {\bf B472} (2000) 382, 
hep-ph/9907532.

\bibitem{bizarre}  G.~Altarelli and F.~Feruglio, JHEP {\bf 11} (1998) 021, hep-ph/9809596; 
Phys.\ Lett.\ {\bf B451} (1999) 388, hep-ph/9812475;
Q. Shafi and Z. Tavartkiladze, Phys. Lett. {\bf B482} (2000) 145, hep-ph/0002150;
G.~Altarelli, F.~Feruglio and I.~Masina, JHEP {\bf 0011} (2000) 040, hep-ph/0007254;
S. F. King and M. Oliveira, Phys. Rev. {\bf D63} (2001) 095004, hep-ph/0009287; 
S. F. King and G. G. Ross,  Phys.Lett. {\bf B520} (2001) 243, hep-ph/0108112.

\bibitem{u2} R. Barbieri, P. Creminelli and A. Romanino, Nucl. Phys. {\bf B559} (1999) 17, hep-ph/9903460;
D. Falcone, Phys. Rev. {\bf D64} (2001) 117302, hep-ph/0106286; 
A. Aranda, Phys. Rev. {\bf D65} (2002) 013011, hep-ph/0109120.

\bibitem{prev} E. Kh. Akhmedov, Phys. Lett. {\bf B467} (1999) 95, hep-ph/9909217;
S. M. Barr and I. Dorsner, Nucl. Phys. {\bf B585} (2000) 79, hep-ph/0003058; 
I. Dorsner and S. M. Barr, Nucl. Phys. {\bf B617} (2001) 493, hep-ph/0108168.

\bibitem{efd} G. C. Branco and J. I. Silva-Marcos, Phys. Lett. {\bf B526} (2002) 104, hep-ph/0106125.

\bibitem{Casas01} J.~A.~Casas and A.~Ibarra, Nucl. Phys. {\bf B618} (2001) 171, hep-ph/0103065. 

\bibitem{Davidson01}
S.~Davidson and A.~Ibarra, JHEP {\bf 0109} (2001) 013, hep-ph/0104076.
%
\bibitem{athand}
W.~Buchmuller, D.~Delepine and F.~Vissani, Phys.\ Lett.\ {\bf B459} (1999) 171;
K.~S.~Babu, B.~Dutta and R.~N.~Mohapatra, Phys.\ Lett.\ {\bf B458} (1999) 93;
W.~Buchmuller, D.~Delepine and L.~T.~Handoko,
Nucl.\ Phys.\ {\bf B576} (2000) 445;
J.~Ellis, M.~E.~Gomez, G.~K.~Leontaris, S.~Lola and D.~V.~Nanopoulos,
Eur.\ Phys.\ J.\ {\bf C14} (2000) 319; 
J. L. Feng, Y. Nir and Y. Shadmi, Phys. Rev. {\bf D61} (2000) 113005;
J.~Hisano and K.~Tobe, Phys.\ Lett.\ {\bf B510} (2001) 197;
D.~F.~Carvalho, J.~Ellis, M.~E.~Gomez and S.~Lola, Phys. Lett. {\bf B515} (2001) 323, hep-ph/0103256;
T.~Blazek and S.~F.~King, Phys. Lett. {\bf B518} (2001) 109, hep-ph/0105005;
J.~Sato, K.~Tobe and T.~Yanagida, Phys.\ Lett.\ {\bf B498} (2001) 189;
J.~Sato and K.~Tobe, Phys.\ Rev.\ {\bf D63} (2001) 116010;
A. Kageyama, S. Kaneko, N. Shimoyama and M. Tanimoto, Phys. Lett. {\bf B527} (2002) 206, hep-ph/0110283;
hep-ph/0112359.

\bibitem{lms1}  S. Lavignac, I. Masina and C. A. Savoy, Phys. Lett. {\bf B520} (2001) 269, hep-ph/0106245.

\bibitem{sens} J. Ellis, J. Hisano, S. Lola and M. Raidal, Nucl.Phys. {\bf B621} (2002) 208, hep-ph/0109125.

\bibitem{more}  S. Lola and G. G. Ross, Nucl. Phys. {\bf B553} (1999) 81, hep-ph/9902283;
Z. Berezhiani and A. Rossi, Nucl. Phys. {\bf B594} (2001) 113, hep-ph/0003084; 
M. Abud and F. Buccella, Int. J. Mod. Phys. {\bf A16} (2001) 609, hep-ph/0006029;
B. Stech, Phys. Rev. {\bf D62} (2000) 093019, hep-ph/0006076;
C. H. Albright and S. M. Barr, Phys. Rev. {\bf D64} (2001) 073010, hep-ph/0104294; 
C. H. Albright and S. Geer, hep-ph/0108070;
K. S. Babu and S. M. Barr, Phys. Lett. {\bf B525} (2002) 289, hep-ph/0111215;
N. N. Singh and M. Patgiri, hep-ph/0112123;
K. S. Babu and R. N. Mohapatra, hep-ph/0201176;
C. D. Froggatt, H. B. Nielsen and Y. Takanishi, hep-ph/0201152.

\bibitem{RGE_nu} J.~Ellis and S.~Lola, Phys.\ Lett.\ {\bf B458} (1999) 310;
J.~A.~Casas, J.~R.~Espinosa, A.~Ibarra and I.~Navarro, Nucl.\ Phys.\ {\bf B569} (2000) 82;
P.~H.~Chankowski, W.~Krolikowski and S.~Pokorski, Phys.\ Lett.\ {\bf B473} (2000) 109;
R. Barbieri, G. Ross and A. Strumia, JHEP {\bf 9910} (1999) 020, hep-ph/9906470;
P. H. Chankowski and S. Pokorski, hep-ph/0110249.

\bibitem{Smir93} A. Yu. Smirnov, Phys. Rev. {\bf D48} (1993) 3264, hep-ph/9304205. 

\bibitem{HKlept} M. Hirsch and S. F. King, Phys. Rev. {\bf D64} (2001) 113005, hep-ph/0107014.

\bibitem{pet82} S. T. Petcov,  Phys. Lett. {\bf B110} (1982) 245.

\bibitem{barb98} R. Barbieri, L. J. Hall, D. Smith, A. Strumia and N. Weiner, JHEP {\bf 9812} (1998) 017,
hep-ph/9807235.

\bibitem{iHcase} Y. Nir, JHEP {\bf 0006} (2000) 039, hep-ph/0002168.

\bibitem{iHbreak} S. F. King and N. N. Singh, Nucl. Phys. {\bf B596} (2001) 81, hep-ph/0007243;
L. Lavoura and W. Grimus, JHEP {\bf 0009} (2000) 007, hep-ph/0008020; W. Grimus and L. Lavoura,
JHEP {\bf 0107} (2001) 045, hep-ph/0105212.

\bibitem{cpbranco} J. A. Aguilar-Saavedra and G. C. Branco, Phys. Rev. {\bf D62} (2000) 096009, hep-ph/0007025;
H. Fritzsch and Z. Xing, Phys. Lett. {\bf B517} (2001) 363, hep-ph/0103242.

\bibitem{FN} C.~D.~Froggatt and H.~B.~Nielsen,
Nucl.\ Phys.\ {\bf B147} (1979) 277.

\bibitem{dedmu}  A. Romanino and A. Strumia, Nucl. Phys. {\bf B622} (2002) 73, hep-ph/0108275;
J. Ellis, J. Hisano, M. Raidal and Y. Shimizu, Phys. Lett. {\bf B528} (2002) 86, hep-ph/0111324.


\bibitem{Gonz01} M.~C.~Gonzalez-Garcia, M.~Maltoni, C.~Pena-Garay
and J.~W.~F.~Valle, Phys.\ Rev.\ {\bf D63} (2001) 033005, hep-ph/0009350.

\bibitem{bach} J. N. Bahcall, M. C. Gonzalez-Garcia and C. Pena-Garay, hep-ph/0111150.
 


\bibitem{Borzumati86} F.~Borzumati and A.~Masiero,
Phys.\ Rev.\ Lett.\ {\bf 57} (1986) 961.

\bibitem{Barbieri95} R.~Barbieri, L.~Hall and A.~Strumia,
Nucl.\ Phys.\ {\bf B445} (1995) 219.
%
\bibitem{Hinchliffe01}
I.~Hinchliffe and F.~E.~Paige, Phys.\ Rev.\ {\bf D63} (2001) 115006.
%
\bibitem{Hisano99} J.~Hisano, T.~Moroi, K.~Tobe and  M.~Yamaguchi,
Phys.\ Rev.\ {\bf D53} (1996) 2442;
J.~Hisano and D.~Nomura, Phys.\ Rev.\ {\bf D59} (1999) 116005.
%

\bibitem{futdirexp}
L.~M.~Barkov {\it et al.}, proposal for an experiment at PSI,
http://meg.psi.ch/.
%



\bibitem{fsv}  F. Feruglio, A. Strumia and F. Vissani, hep-ph/0201291.



 
\end{thebibliography}
\end{document}